\documentclass[12pt]{iopartA}
\usepackage{epsfig,float}
\usepackage{epstopdf}
\usepackage{graphicx}
\usepackage{iopams}

\makeatletter

\newcommand{\beq}{\begin{equation}}
\newcommand{\eeq}{\end{equation}}

\eqnobysec

\newcommand{\appsection}{\addtocounter{section}{1} \setcounter{equation}{0}
                         \section*{Appendix \Alph{section}}}

\newcommand{\intd}{\mathrm{d}}

\newcommand{\ZZ}{\mathbb{Z}}

\newcommand{\ket}[1]{|#1 \rangle}
\newcommand{\braket}[2]{\langle #1 | #2 \rangle}

\newcommand{\brakettt}[3]{\langle #1 | #2 |#3 \rangle}

\usepackage{latexsym}

\makeatother

\begin{document}

\title[The particle spectrum of the TIM with spin reversal symmetric
  perturbations]{The particle spectrum of the Tricritical Ising Model with
  spin reversal symmetric perturbations}

\author{
L Lepori$^{1,2}$,
G Mussardo$^{1,2,3}$,
G Zs T\'oth$^{1,2,4}$
}

\address{$^1$ International School for Advanced Studies (SISSA),
Via Beirut 2-4, 34014 Trieste, Italy}
\address{$^2$ INFN Sezione di Trieste, Italy}
\address{$^3$ The Abdus Salam International Centre for Theoretical Physics,
  Trieste, Italy}
\address{$^4$ Research Institute for Particle and Nuclear Physics,
Hungarian Academy of Sciences, Pf.\ 49, 1525 Budapest,
Hungary}

\eads{\mailto{lepori@sissa.it}, \mailto{mussardo@sissa.it},
  \mailto{toth@sissa.it} and \mailto{tgzs@cs.elte.hu}}

\begin{abstract}
We analyze the evolution of the particle spectrum of the Tricritical Ising
Model by varying the couplings
 of the energy and vacancy density fields. The particle content changes from
 the spectrum of a supersymmetric theory (either of an exact or a
 spontaneously broken supersymmetric theory) to the spectrum of
 seven particles related to the underlying $E_7$ structure.  In the
 low temperature phase some
 of these excitations are topologically charged particles that are stable
 under an arbitrary
variation of the parameters. The high and low temperature phases
of the model are related by duality. In some regions of the two
couplings there are also present false vacua and sequences of bound states. In order
to study the non-integrable features of this model we employ the
Form Factor
 Perturbation Theory and the Truncated Conformal Space Approach.
\end{abstract}

\begin{center}
{This paper is dedicated to the memory of
Aliosha Zamolodchikov.}
\end{center}

\maketitle

\section{Introduction}
\label{sec.1}

The two-dimensional Tricritical Ising Model (TIM), which is the second in the
unitary series of
conformal minimal models after the critical Ising model \cite{BPZ}, describes
critical phenomena in a variety of systems with tricritical points. It also
exhibits superconformal symmetry, being the first member
 of the series of superconformal minimal models \cite{FQS}, and it has
 interesting symmetries related to the coset constructions
 $su(2)_2\otimes su(2)_1/su(2)_3$ and $e(7)_1\otimes e(7)_1/e(7)$
 \cite{Goddard1986, Lukyanov1989}.
 Due to these properties the TIM has attracted interest for many years.

In the present paper we study the particle spectrum of the TIM perturbed by
 those relevant scaling
 fields which are invariant under spin reversal. These fields are the energy
 density $\varepsilon$
and the vacancy density $t$, with conformal weights given by $\Delta_{2} =
 1/10$ and $\Delta_{4} = 3/5$
respectively\footnote[1]{For further details on the model and its symmetry we
 refer the reader to
 the article \cite{LMC}.}. The action of the perturbed model can  formally be written as
\beq
\label{eq.action}
\mathcal{A}=\mathcal{A}_{0}+g_2\int \varepsilon(z,\bar{z})\, \intd^2 z
+g_4\int \,t(z,\bar{z})\, \intd^2 z,
\eeq
where $\mathcal{A}_{0}$ is the action of the TIM at its conformal point, whereas $g_2$ and $g_4$ are the two coupling constants of our theory. Their dimensions are fixed by the conformal weights of the
conjugate fields, $g_i \sim {\cal M}^{2 - 2 \Delta_i}$,
where ${\cal M}$ is an arbitrary mass scale of the off-critical model. The ratios of the particle masses of the model (\ref{eq.action}), being universal quantities, depend only on a single dimensionless combination $\eta$ of the two coupling constants $g_2$ and $g_4$, that we choose as
\beq
\eta=\frac{g_4}{|g_2|^{\frac{2-2\Delta_4}{2-2\Delta_2}}}=
\frac{g_4}{|g_2|^{\frac{4}{9}}}\ .
\eeq
In the following we write $\eta_+$ if $g_2>0$ and
$\eta_-$  if $g_2<0$.

The evolution of the particle spectrum by varying $\eta$ is an important
characteristic of the
class of universality associated with the TIM. It is worth stressing that the
action (\ref{eq.action}) provides the simplest example of a bosonic theory
having kinks in its spectrum which do not get confined under continuous variations
of the coupling constants. To appreciate this feature, one has to recall that,
in a generic bosonic theory, the kinks are quite fragile objects: they get
 generally confined under a small perturbation of the parameters  (see, for
 instance, refs.\ \cite{DMS,DM}). The stability of the kinks can be guaranteed,
 for instance, by supersymmetry but in theories that have also fermions
 \cite{GMsusy}. In purely bosonic theories, the robustness of the kink states may only come via a fine-tuning mechanism or special symmetry of the perturbing operators. In
the model analyzed in this paper, the latter property is the spin reversal invariance of both perturbations. The study of the particle spectrum of this model is, in
any case, interesting in itself, since the action (\ref{eq.action}) is
non-integrable if both $g_2$ and $g_4$ are non-zero.

\section{A quick glimpse into the model}

In the TIM, the parameter $g_2$ can be interpreted as the difference between the
actual temperature and its critical value. The $g_2=0$ critical line divides the
phase plane into the high temperature ($g_2 > 0$) and the low temperature
($g_2 < 0$) halves, which are mapped into each other by a duality
transformation under which $(g_2,g_4) \leftrightarrow (-g_2,g_4)$.
Along the critical line $g_2 =0$, the model (\ref{eq.action}) is integrable
and has also supersymmetry \cite{ZZ,Z2,KMS}:  the $g_4 > 0$ half line consists
of a massless renormalization group flow to the Ising critical point in which the supersymmetry is spontaneously
broken, while the $g_4<0$ half line consists of first order phase
 transition points with unbroken supersymmetry. The meeting point of these
 half lines is the tricritical point.
The model is also integrable along the $g_4=0$ line where its particle spectrum
and
 S-matrix are related to the $E_7$ Lie-algebra \cite{ChM,FaZam}, therefore we
 shall refer to this
 line as the  $E_7$-related line.

\begin{figure}[H]
\begin{indented}
\item
\includegraphics[
  scale=0.6]{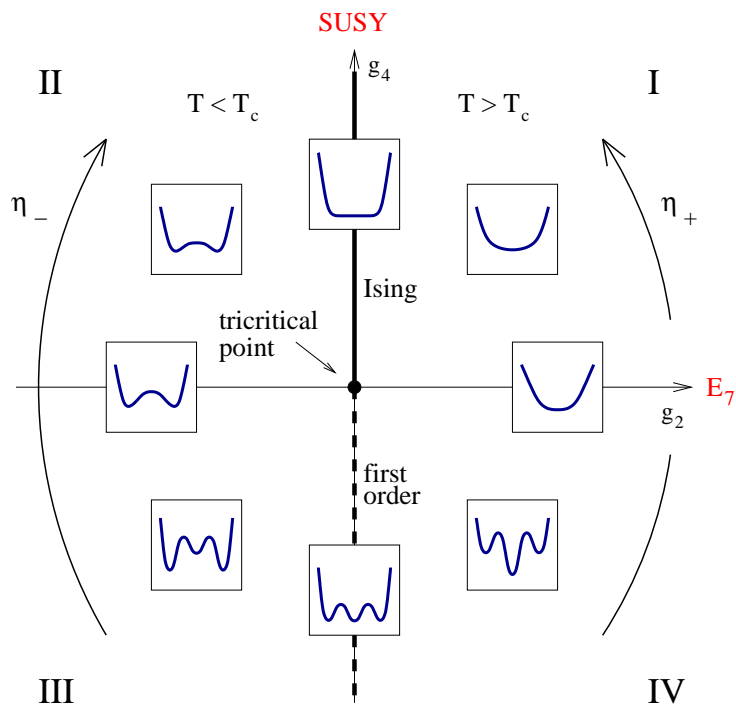}
\end{indented}
\caption{\label{fig.diag}
{\em Landau-Ginzburg potentials associated with the various parts of the phase space.}
}
\end{figure}

It is well known that the $\phi^6$ Landau-Ginzburg theory has a tricritical
point which is
described by the TIM \cite{ZamLG}, in particular the model (\ref{eq.action}) admits another description in terms of the action 
\beq
\label{eq.lg1}
\mathcal{A}_{LG}=\int \intd^2 x\, \left( \frac{1}{2}(\partial_\mu
\phi)^2+\phi^6+a_2\phi^2+
a_4\phi^4\right),
\end{equation}
with the identification 
$\varepsilon \sim\ : \hspace{-1.6mm}   \phi^2  \hspace{-1.6mm} :$,  $t \sim\
 :  \hspace{-1.6mm}     \phi^4  \hspace{-1.6mm}  :$. In
this formulation the
tricritical point corresponds to $a_2 = a_4=0$; the spin reversal
transformation acts on $\phi$ as $\phi\mapsto - \phi$. This correspondence
allows one to associate the Landau-Ginzburg potentials with the various
 parts of the $(g_2,g_4)$ phase plane, as shown in Figure \ref{fig.diag}.

The shape of the potentials in Figure \ref{fig.diag} provides a useful guide
to the nature of the excitations at the various points of the $(g_2,g_4)$
plane. For instance, in the high temperature phase,  where the
spin reversal symmetry is unbroken, the potential has a unique vacuum and
therefore the particles do not have topological charge in this phase. Moving
up from the positive horizontal axis in the first quadrant, the curvature of
the minimum decreases until it vanishes when we reach the positive vertical
axis: the theory has a
 massless spectrum at this point. Instead, moving down from the positive
 horizontal axis in the forth quadrant, two metastable vacua start to appear
 and they become abruptly degenerate with the true vacuum once the negative
 vertical axis is reached. Right at this point the theory has kink excitations,
 i.e.\ topologically charged
particles, that interpolate between the three vacua.

In the low temperature phase, i.e.\ in the second and third quadrants, the
 spin reversal symmetry is instead spontaneously broken and the ground state
 is doubly degenerate,  hence in these regions we expect that the spectrum
 consists of kinks and possible bound states  thereof. Moreover, it is
 sufficient to look at the
 evolution of the potential in these quadrants in order to conclude that the
 kinks 
of this model are always stable excitations (at least those with
 lowest mass), no matter how we vary the coupling constants.  In more detail,
 starting from the negative horizontal axis and moving up in the third
 quadrant, the barrier between the two vacua decreases until it disappears
 when we reach the positive vertical axis. Hence, we expect the masses of
 the kinks and their possible bound states to decrease along this trajectory so
 that they adiabatically vanish once we arrive at the positive vertical
 axis. On the other hand, moving down from the negative horizontal axis in the
 third quadrant, a metastable vacuum starts to appear in the middle of the
 barrier: this state
 becomes abruptly degenerate with the two existing vacua just when we reach
 the negative vertical axis. At this point, the kink excitations between the
 two original vacua break down and give rise to two
 different sets of topological excitations relative to the new threefold vacuum structure.

In the following we are going to support the physical scenario described
 above by using a combination of analytic and numerical methods. The first method is
the Form Factor Perturbation Theory (FFPT) \cite{DMS} that allows us to
 get  analytic information on the spectrum in the vicinity of the integrable lines. The second method is
the Truncated Conformal Space Approach (TCSA) \cite{YZ}, which is one of the
best suited numerical approaches for extracting the spectrum of
perturbed conformal field theories. The
spin reversal and the duality symmetries of the TIM play also an important
role in the implementation of both methods.
Let's finally mention that the two integrable cases $g_2=0$ and $g_4=0$ were
 studied by TCSA in \cite{LMC}; the present work can thus be  regarded as an
 extension of those studies to the
 non-integrable ($g_2\ne 0$, $g_4\ne 0$) domain.

\section{The two methods}
In this section we briefly discuss the two techniques
that have been employed to study the evolution of the spectrum of our model.

\subsection{Form Factor Perturbation Theory}

The Form Factor Perturbation Theory permits to investigate the behaviour of a
 theory in the vicinity of an integrable direction, either if  the integrable
 theory is massive \cite{DMS,DM} or massless \cite{CM}. Let
 $\mathcal{A}_{\rm{int}}$ be the action of the integrable model and $\Upsilon(x)$
 the field that moves the system
 away from integrability, so that nearby the action of the theory can be written as
\begin{equation}
{\cal A}\,=\,{\cal A}_{\rm{int}} + \lambda\,\int \,\intd^2 x\,\,\Upsilon(x)\ .
\label{perturbedaction}
\end{equation}
If the original integrable model has scalar
excitations\footnote[2]{The rapidity variable $\theta$ expresses the dispersion
  relation of the excitations. If we consider a massive particle of mass $m$,
  its energy and momentum are given by $E = m \cosh\theta$, $p = m
  \sinh\theta$. If, instead, the particle is massless, it can be a right or a left
  mover excitation.  The dispersion relations of a right mover are $E_{R} = p =
  M \rme^{\theta}$, while those of a left mover $E_L = -p = M \rme^{-\theta}$, where $M$
  is an arbitrary mass scale.} $A_1(\theta),A_2(\theta),\ldots$, the first
order correction
 in $\lambda$ to their mass $M_k$ is given by
\begin{equation}
\delta M_{k}^2 \,\simeq \,2\,\lambda \,F_{k}^{\Upsilon}(\rmi\pi)\,\,,
\label{deltam1}
\end{equation}
where
\begin{equation}
F_{k}^{\Upsilon}(\theta_{12}) \,\equiv\,\brakettt{0}{\Upsilon(0)}{A_k(\theta_1)
\,A_{k}(\theta_2)}
\label{aafar}
\end{equation}
is the two-particle form factor of the operator $\Upsilon(x)$, with $\theta_{12} = \theta_1 - \theta_2$.
A similar formula also holds if the integrable model has massless excitations,
with a proper interpretation of the rapidity variable $\theta$ \cite{CM}. If, instead,
the integrable theory has topological kinks $\ket{K_{ab}(\theta)}$
interpolating between the vacuum states $\ket{a}$ and $\ket{b}$, the relevant formula is
\begin{equation}
\delta M_{ab}^2 \,\simeq \,2\,\lambda \,F_{ab}^{\Upsilon}(\rmi\pi)\,\,,
\label{deltam2}
\end{equation}
where
\begin{equation}
F_{ab}^{\Upsilon}(\theta_{12}) \,\equiv\,_k\brakettt{0}{\Upsilon(0)}{K_{ab}(\theta_1)
\,K_{ba}(\theta_2)}\ .
\label{aabar}
\end{equation}

In integrable field theories the form factors of a generic scalar operator
 $\Upsilon(x)$ can be computed exactly \cite{KW,Smirnov} and therefore we can
 study the evolution of the spectrum near to the integrable lines. However, one
 has to be careful when applying the formulas (\ref{aafar}) and (\ref{aabar})
 since
 the two-particle form factor can be singular at $\theta_{12} = \rmi \pi$. This happens if the perturbing operator $\Upsilon(x)$ has
a semi-local index $\gamma$ with respect to the particle excitations (this is
usually the
 case of the topological kinks, for instance). In this case the point $\theta_{12} =\pm \rmi\pi$ is the location of a simple pole of the two-particle form factor, whose residue is given by \cite{DM,Smirnov}
\begin{equation}
-\rmi\,{\rm{Res}}_{\,\theta_{12}=\pm \rmi\pi}\,F_{ab}^{\Upsilon}(\theta_{12})\,=\,
(1-\rme^{\mp 2\rmi\pi\gamma})\,\,_k\brakettt{0}{{\Upsilon}}{0}_k
\ .
\label{pole}
\end{equation}
The result of the brief analysis given above can be summarized as follows.
 If $F_l^{\Upsilon}(\rmi\pi)$ is finite, the corresponding particle or kink
 excitation survives the perturbation and its mass is adiabatically shifted
 from the original value. Vice versa, if $F_l^{\Upsilon}(\rmi \pi)$ is divergent,
 the corresponding excitation no longer survives as asymptotic particle of the
 perturbed theory, i.e.\ it gets confined.

\subsection{The truncation method}
\label{sec.3}

The Truncated Conformal Space Approach (TCSA) \cite{YZ} can be used to study
perturbed
conformal field theories. It consists in a finite size analysis of the
spectrum of a theory on a cylinder of circumference $R$. This is realized by
truncating the conformal basis at a certain level
 $n$ in the Verma modules of the irreducible representations of the primary
fields and
 solving the eigenvalue and eigenvector problem of the truncated Hamiltonian operator numerically.

To extract a quantity $F$ relative to the $R\to\infty$ limiting model, one
 calculates $F(R,n)$ (i.e.\ $F$ at a cylinder circumference $R$ and truncation
 level $n$) for several values of $R$, and possibly also for several values of
 $n$. Usually there is a range $R_{F,\rm{min}} < R <
 R_{F,\rm{max}}(n)$ such that in this range  $F(R,n)$  is an approximately constant
 function of $R$, and its value can be regarded as a good
 approximation of the infinite-volume value of $F$. This range is called the
 {\em physical window} for $F$.
 Both $R_{F,\rm{min}}$ and $R_{F,\rm{max}}(n)$ depend on
$F$, and   $R_{F,\rm{max}}(n)$ also depends on $n$ and is
expected to tend to infinity as $n$ is increased.
Hence, the approximation to $F$ usually improves if $n$ is increased. For $R >
 R_{F,\rm{max}}(n)$ the truncation effects spoil the approximation to $F$ and give
 rise to an unphysical window of values for this
 quantity.

We determine the masses of the particles from
the energy differences  
\beq
\Delta E_i(R,n) \equiv (E_i(R,n)- E_0(R,n))\ ,  
\eeq
where $E_0(R,n)$ is the lowest energy line. In our case, the individual energy levels  $E_i(R,n)$ diverge as $n \to \infty$  if $g_4 \ne 0$, since the conformal weight of the perturbation $t$ is greater than $1/2$, the energy differences  $E_i(R,n)- E_0(R,n)$ are however convergent \cite{KlM}.
We also use the  TCSA to calculate form factors and vacuum expectation
values. This can be done by computing numerically the eigenvectors
belonging to the various energy eigenvalues.

The Hamiltonian operator of the model (\ref{eq.action}) on a cylinder of circumference $R$ is
\begin{eqnarray}
\label{eq.Ham}
\fl
H=\frac{2\pi}{R}\left(L_0+\bar{L}_0-\frac{c}{12}\right) \nonumber\\ 
+
2\pi g_2\lambda_2 R^{-2\Delta_2}\int_0^R \varepsilon(x,0)\, \intd
x+
2\pi g_4\lambda_4 R^{-2\Delta_4}\int_0^R t(x,0)\, \intd
x,
\end{eqnarray}
where $c=7/10$ is the central charge of the TIM, whereas $L_0$ and $\bar{L}_0$
 are the zero
 index generators of the chiral Virasoro algebras.
For convenience, we have chosen to introduce the quantities
\beq
\lambda_2= (2\pi)^{2\Delta_2-1}\times  0.0928344\ ,\qquad
\lambda_4= (2\pi)^{2\Delta_4-1}\times  0.1486960\  .
\eeq
These are precisely the values of the coupling constants that ensure a mass gap $M=1$ when
we specialize the above Hamiltonian to the  integrable massive cases ($g_2 =\pm 1, g_4 =0)$ and ($g_2 =0, g_4 = -1)$ \cite{Fateevcoupling}.
Further data which are necessary to implement the TCSA for the TIM can be found in \cite{LMC}.

To study the mass spectrum it is sufficient to consider the zero momentum
 subspace only.
 Moreover, one can also treat the  parts of the
 Hamiltonian which are even or odd with respect to spin reversal  separately\footnote[3]{In addition to the two spin reversal
 even fields $\varepsilon$ and $t$, the TIM possesses
other two even fields, the identity operator $I$ and the supersymmetry
 generator $G$, with conformal weights $0$ and $3/2$ respectively.
 The spin reversal odd sector is generated by the magnetization operator
 $\sigma$ and the
sub-leading magnetization $\sigma'$, with $\Delta_{\sigma}=3/80$ and $\Delta_{{\sigma}'} = 7/16$.}
. 
In our
 calculation we used periodic boundary conditions, which selects the sector
of zero topological charge
 in the Hilbert space of the multi-particle states. As we will
 show below, there may be other phenomena related to the boundary conditions
 and to the finite size of the system that have to be properly
 taken into account in order to interpret the spectrum correctly.

\begin{figure}[H]
\begin{center}
\includegraphics[
  scale=0.6]{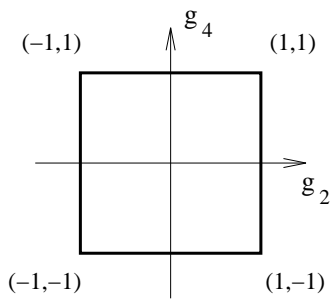}
\end{center}
\caption{\label{fig.diag2}
{\em The square in the $(g_2,g_4)$ plane along which the TCSA calculations were done.}
}
\end{figure}

In our numerical calculations we varied the coupling constants $g_2$ and
 $g_4$ along
 the square with corners $(\pm 1,\pm 1)$ shown
in Figure \ref{fig.diag2}. The truncation level that we used is $n=8$,
 that corresponds
 to a Hamiltonian truncated to $1624$ conformal states. The finite volume spectra shown in Figures
\ref{3vacTCSA}, \ref{decay}, \ref{h1} and \ref{h2}  are always 
plotted as functions of $mR$, where $m$ is the mass gap, whose value depends
 however on the region of the coupling constants  which each figure refers to.

\section{The spectrum in the high temperature phase}
\label{sec.2}

In this section we describe our results on the particle spectrum of the model
 (\ref{eq.action}) in the high temperature phase, made up of the first and the
 fourth quadrant of the $(g_2,g_4)$ plane.
 This phase is described by the variable $\eta_+ \in (-\infty,+\infty)$:
the value $\eta_+ = -\infty$ corresponds to the negative vertical axis,
 $\eta_+=0$ to the positive
 horizontal axis, and $\eta_+=+\infty$ to the positive vertical axis.

In this phase the $\ZZ_2$ spin reversal symmetry is exact and there is a unique vacuum of the theory
(although in the fourth quadrant there are also two false vacua). Hence here the excitations are
 ordinary scalar particles.  In order to follow the evolution of the spectrum it is convenient to start our
analysis from the vicinity of the negative vertical axis, i.e.\ from the fourth quadrant.

\subsection{The fourth quadrant}
\label{sec.fourth}

\begin{figure}
\begin{indented}
\item
\includegraphics[
 scale=0.5]{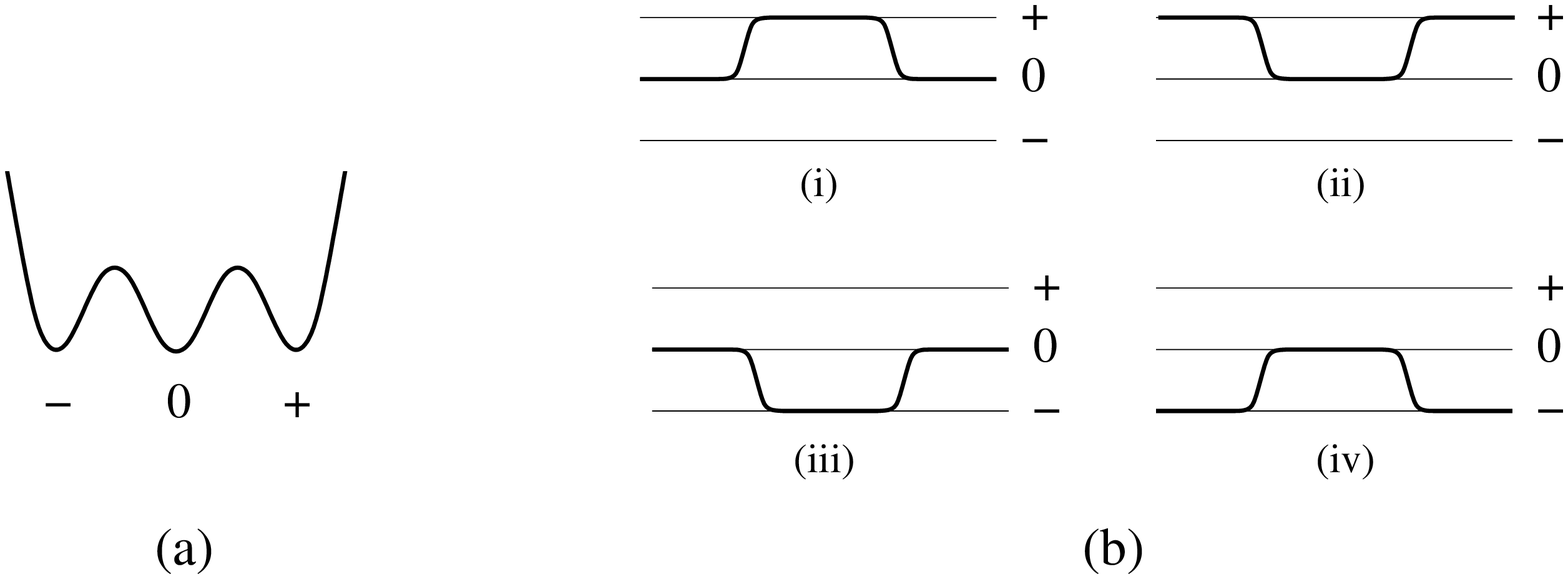}
\end{indented}
\caption{\label{3vacua}{\em
(a) The three vacua in the first-order transition point. (b) Unbound kink-antikink pairs. On the cylinder with periodic boundary the configurations (i) and (ii)
  represent the same state, and so do (iii) and (iv).}
}
\end{figure}

Precisely at $\eta_+ = -\infty$, the model has a first order phase
 transition. This means that along the negative vertical axis the theory has
 three degenerate ground states $\ket{0}$, $\ket{+}$ and $\ket{-}$:
 the vacuum $\ket{0}$ is even under the
spin reversal symmetry $Q$, while $\ket{+}$ and $\ket{-}$ are mapped into each
other:
 $\ket{-} = Q\ket{+}$ (see Figure \ref{3vacua}.a). Right at $\eta_+ =
 -\infty$, the theory presents
also an exact supersymmetry: the particle spectrum consists of four kinks
 $\ket{K_{-0}}$, $\ket{K_{0-}}$, $\ket{K_{0+}}$ and $\ket{K_{+0}}$ of equal
 mass. These kinks, that do not have bound states,
 provide an irreducible representation of supersymmetry and their exact S-matrix
 was
obtained in \cite{Z2}. The action of the spin reversal symmetry $Q$ on the kinks is given by
$Q\ket{K_{-0}} = \ket{K_{+0}}$, $Q\ket{K_{0-}} = \ket{K_{0+}}$.

\begin{figure}
\begin{indented}
\item
\includegraphics[
 scale=0.45]{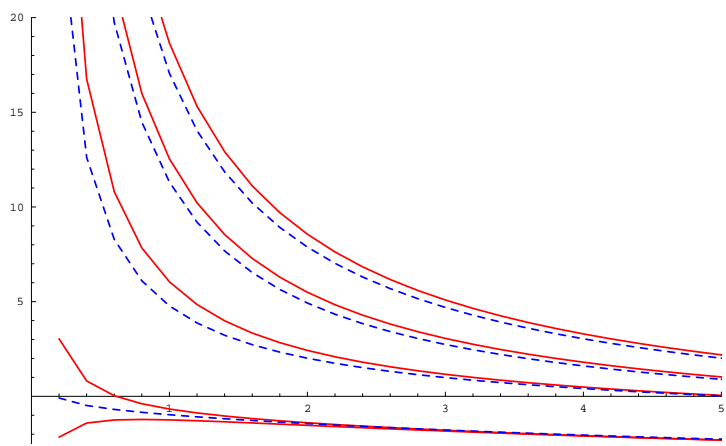}
\end{indented}
\caption{\label{3vacTCSA}{\em
The lowest nine energy levels $E_i$, $i=0...8$ in the first-order transition point as functions
  of $mR$, obtained by TCSA. Red lines are even, blue (dashed) lines are odd
  under spin reversal. $m$ denotes the mass gap.}
}
\end{figure}

On a cylinder of circumference $R$, the three degenerate vacua give rise to three exponentially split energy lines, for the phenomenon of tunneling that occurs at a finite volume. With periodic boundary conditions, there are however no energy lines corresponding to the one-particle states of the kinks. Above the
three exponentially degenerate lines of the ground states, one finds instead
doublets 
of lines corresponding to the neutral kink-antikink configurations (see Figure \ref{3vacTCSA}). The proper counting of the states appearing in the finite volume with periodic boundary conditions is explained in Figure \ref{3vacua}.b.

\begin{figure}[h]\label{linearfigure}
\hskip 30pt
\begin{minipage}[b]{.45\linewidth}
\vspace{5mm}
\centering\psfig{figure=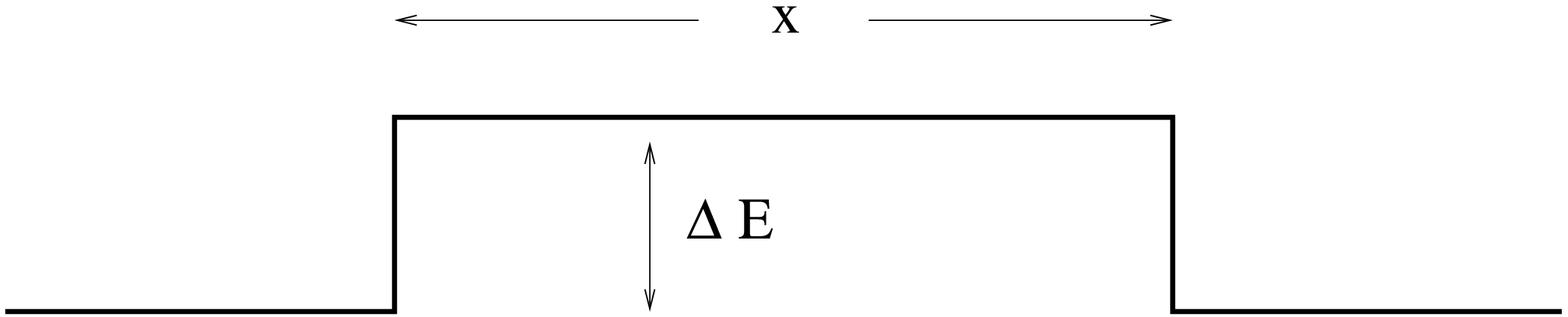,width=\linewidth}
\begin{center}
{\bf (a)}
\end{center}
\end{minipage} \hskip 55pt
\begin{minipage}[b]{.25\linewidth}
\centering\psfig{figure=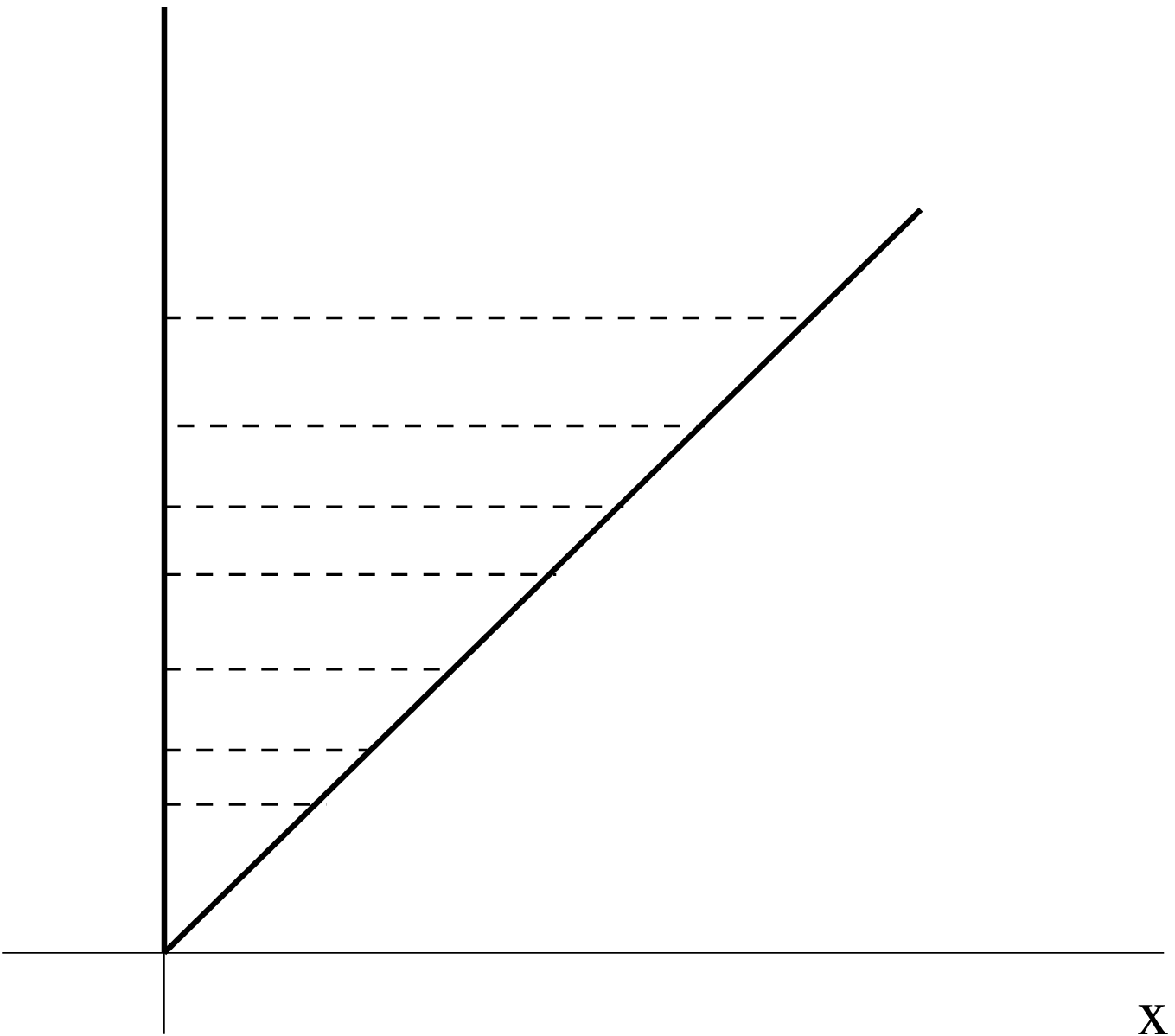,width=\linewidth}
\begin{center}
{\bf (b)}
\end{center}
\end{minipage}
\caption{
\label{AAA}
{\em (a) Kink-antikink state at distance $x$, where $\Delta E$ is the gap of the unbalanced vacua.  The kink and the antikink are subject to a linear confining potential (b), with slope given by $\Delta E$, and they give rise to an infinite sequence of bound states.}}
\end{figure}

It is now important to understand what happens if we deform the theory away
from this integrable situation, moving into the fourth quadrant by means of
the perturbation $g_2 \int \varepsilon(x) \intd x$ ($g_2 > 0$). The drastic effect
of this perturbation is to lift the degeneracy of the three vacua: in the perturbed theory, the central
vacuum $\ket{0}$ becomes the true ground state of the theory, whereas the other
two vacua $\ket{\pm}$  become metastable ground states, separated from $\ket{0}$ by a
gap 
\beq
 \Delta E\sim
g_2(\brakettt{\Omega}{\epsilon}{\Omega}-\brakettt{0}{\epsilon}{0})
\ , 
\eeq
where
$\ket{\Omega} = \ket{\pm}$. As a consequence, we expect that the kinks of the unperturbed
system will get confined as soon as we move away from $\eta_+ = -\infty$ by
switching on the coupling $g_2$ of the energy operator $\epsilon$. This can be directly seen 
by the FFPT, reported in Appendix C: the two-particle form
factors on the kink states of the $\epsilon$ operator have in fact a pole at
$\theta = \rmi \pi$ \cite{D}. The linear confining potential between the
constituents of the  two
pairs of neutral kink-antikink states $\ket{K_{0-} K_{-0}}$ and
$\ket{K_{0+}K_{+0}}$
gives rise then to a dense  sequence of bound states, with the number of bound
states going to infinity as $g_2\to 0$
\cite{DM} (see Figure \ref{AAA}).

\vspace{3mm}

\begin{figure}
\begin{center}
\includegraphics[
  scale=0.45]{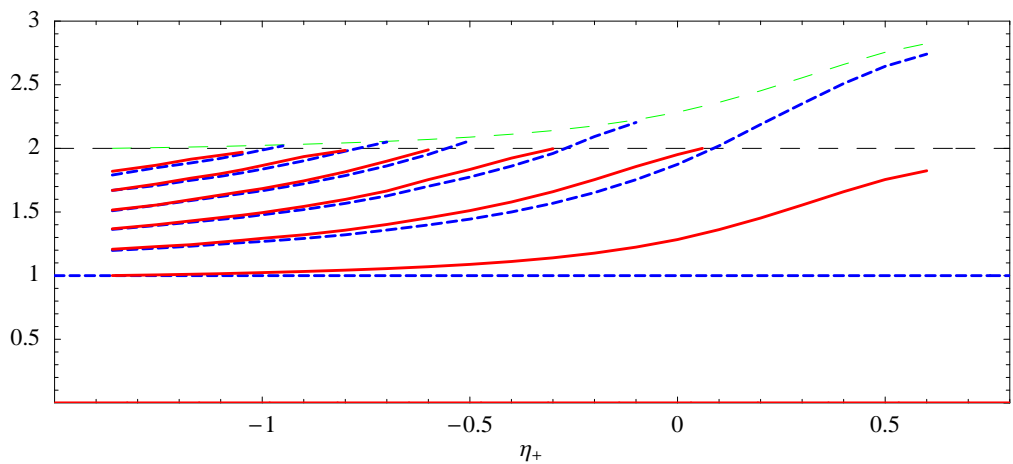}
\end{center}
\caption{\label{fig.ratios}
{\em The particle mass ratios $m_i/m_1$ as functions of
$\eta_+$ in the high temperature phase in the range $-1.361\le \eta_+\le 0.6$, obtained by TCSA.
Even states are shown in red, odd states in blue and with dashed line.
The thresholds $2m_1$ and  $m_1+m_2$ for even and odd particles are drawn with
dashed thin black and green lines.
}}
\end{figure}

This analytic result is indeed confirmed by our numerical calculations. In
particular, Figure  \ref{fig.ratios} shows that the number of
particles below the stability threshold increases to infinity as
the first order transition line is approached (i.e.\ $\eta_+ \to
-\infty$). This scenario is similar to the one observed in the
scaling region of the critical Ising model near the thermal axis
 once it is perturbed by a magnetic field \cite{DMS} and it realizes 
the so-called McCoy-Wu scenario
\cite{McWu}. An important difference from the Ising model,
however, is the $\ZZ_2$ symmetry of our model. At $\eta_+ =
-\infty$, the two-kink states $\ket{K_{0-}K_{-0}}$ and
$\ket{K_{0+}K_{+0}}$ (the rapidities of the kinks are suppressed)
are mapped into each other by spin reversal, so they form
degenerate pairs (although in finite volume this degeneracy is lifted). The two-kink states $\ket{K_{-0}K_{0-}}$ and $\ket{K_{+0}K_{0+}}$ also form similar degenerate pairs, and so do the  two-kink states
$\ket{K_{-0}K_{0+}}$ and $\ket{K_{+0}K_{0-}}$. The degeneracy is lifted as
soon as the perturbation is switched on; 
the arising bound
states are singlets with definite parity. As the kinks do not have
bound states at $\eta_+ = -\infty$, all the bound states in the
neighbourhood $g_2> 0$ of the transition point arise from the free
two-kink configurations mentioned above. Since the $\ket{\pm}$
vacua are lifted, the original two-particle states
$\ket{K_{-0}K_{0-}}$, $\ket{K_{+0}K_{0+}}$, $\ket{K_{-0}K_{0+}}$,
and $\ket{K_{+0}K_{0-}}$ of the continuum  disappear entirely from
the spectrum. In their place, there is a sequence of bound states
$\ket{B_n^\pm}$, which  arises from the even and odd superpositions
of the $\ket{K_{0-}K_{-0}}$ and $\ket{K_{0+}K_{+0}}$ states: 
\beq
\label{eq.bn} \ket{B_n^\pm}\sim
\frac{\ket{K_{0-}K_{-0}}\pm\ket{K_{0+}K_{+0}}}{\sqrt{2}}\  . 
\eeq 
The superscript in $\ket{B_n^\pm}$ denotes parity, the subscript
numbers the elements of the tower of bound states. As we
mentioned, $\ket{B_n^+}$ and
 $\ket{B_n^-}$ are not degenerate. Due to the exactness of the
$\ZZ_2$ symmetry the vacuum is expected to be  even, and then the first state above it, i.e.\
the lowest one-particle state, is expected to be odd, and generally the parity
is expected to alternate
in the sequence of
one-particle states.

The numerical data that we obtained by varying $\eta_+$
indeed confirm that in the high temperature phase the
spin reversal symmetry is unbroken,
there is a unique vacuum and the excitations are singlet
scalar particles with
alternating parity. The disappearance of the two-kink states  $\ket{K_{-0}K_{0-}}$,
$\ket{K_{+0}K_{0+}}$, $\ket{K_{-0}K_{0+}}$ and
$\ket{K_{+0}K_{0-}}$ does not have a drastic effect on the numerical spectrum on the cylinder with periodic
boundary conditions, since the  $\ket{K_{-0}K_{0+}}$, and
$\ket{K_{+0}K_{0-}}$ are filtered out and $\ket{K_{-0}K_{0-}}$ and
$\ket{K_{+0}K_{0+}}$ are identified with $\ket{K_{0-}K_{-0}}$ and
$\ket{K_{0+}K_{+0}}$ under these boundary conditions: what happens is that, energy lines that were previously asymptotic degenerate, now become separated by tiny gaps.

The  doubly degenerate false vacuum can  be seen  in the  $\eta_+ < 0$
 domain  in the TCSA
spectra in the form of a  linearly rising double line-like pattern
 (see Figure \ref{h1}.a).
The slope of the pattern, which equals to the gap $\Delta E$ between the
 false and true vacua, increases from $0$ as  $\eta_+$ is increased, and
 finally the pattern disappears from the finite volume spectrum as $\eta_+$
 approaches $0$.
$\Delta E$ can be calculated by FFPT around the first order phase  transition
 line. At $g_4 = -1$ the exact result to first order in perturbation theory is
\beq 
\frac{\Delta E}{g_2}=0.3076...=2 \Lambda_2\times
(2\pi)^{\frac{1-\Delta_2-\Delta_4}{1-\Delta_4}}
\lambda_4^{\frac{\Delta_2}{1-\Delta_4}} \lambda_2\ , 
\eeq 
where
\beq \label{eq.B2} \Lambda_2 =2.668319... 
\eeq 
is the exact value
of the vacuum expectation values of $\varepsilon$ given in
\cite{FLZZ,FMS},
 and the factor next to $2\Lambda_2$ arises from various
normalization factors
used in the present paper and in \cite{FLZZ,FMS} (see Appendix B for more details on the calculation of
 $\Delta E$). On the other hand, by TCSA we obtain
\beq
\label{eq.deltae}
\frac{\Delta E}{g_2}=0.29,
\eeq
which is in reasonable agreement with the exact value.

\begin{figure}
\begin{center}
\includegraphics[
  scale=0.5]{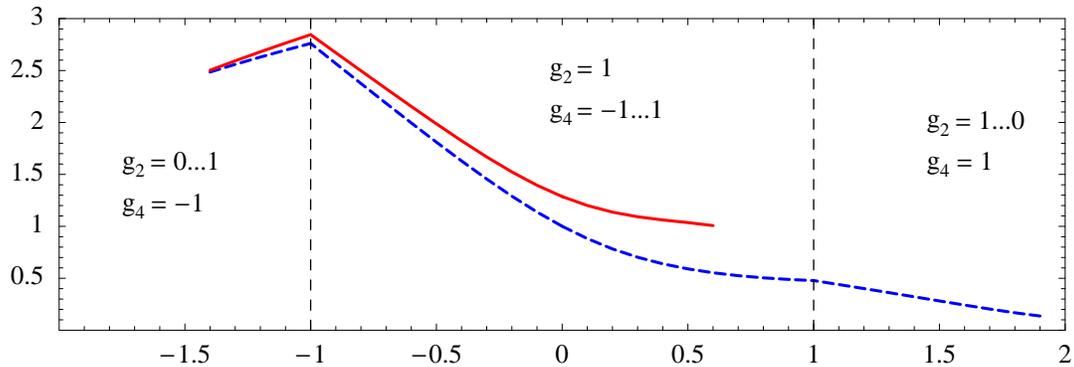}
\end{center}
\caption{\label{fig.absmass}
{\em The particle masses  $m_1$ (in red) and $m_2$  (in blue and with dashed line)
obtained by TCSA along the square shown in Figure \ref{fig.diag2}. The section
$0< g_2<0.6$, $g_4=-1$ is not shown.}
}
\end{figure}

Figure \ref{fig.ratios} shows the numerically calculated mass ratios $m_i/m_1$
($m_1$ being the mass of the lightest particle)
as functions of $\eta_+$. Truncation errors tend to increase for large
absolute values of $\eta_+$, therefore the figure shows only a limited range
around $\eta_+=0$. The
 masses $m_1$ and $m_2$ of the two lightest particles are also shown in Figure \ref{fig.absmass}. The stability threshold, taking into account spin reversal symmetry, is $2m_1$ for even particles and $m_1+m_2$  for odd particles (if there are no even particles, then the threshold is $3m_1$).

Hence, increasing $\eta_+$, the number of stable excitations decreases. Around
 $\eta_+ =0$, i.e.\ near to the positive horizontal axis,
 there are four stable excitations. However, precisely at $\eta_+=0$, the
 theory acquires three more stable particles, all of them above threshold. In
 order to understand the nature of the spectrum near to $\eta_+ = 0$, it is
 obviously more convenient to switch to the integrable formulation of the
 theory along the positive horizontal axis.

\subsection{The $E_7$-related line and FFPT}
\label{sec.6}

As we mentioned in the introduction, the $\eta = 0$ line corresponds to an
 integrable model related to the Toda field theory based on the exceptional
 Lie algebra $E_7$. In the high temperature phase, this integrable model
 has a spectrum composed of seven scalar particles $A_i$, $i=1\dots 7$.
 Their masses $m_i$, $i=1\dots 7$ (see Table \ref{tab.E7}) and S-matrix amplitudes are known exactly
\cite{ChM,FaZam}.

\begin{table}[t]
\caption{\label{tab.E7}{\em Masses and spin reversal parities of the particles on
  the $E_7$-related axis in the high temperature phase.}
}
\begin{indented}
\item
\vspace{2mm}
\begin{tabular}{@{}|l|l|c|}
\hline
mass & $m_i/m_1$ & $\ZZ_2$ parity  \\
\hline
$m_1$ & $1$ & $-$  \\
$m_2=2m_1\cos(5\pi/18)$ & $1.28557$ & $+$    \\
$m_3=2m_1\cos(2\pi/18)$ & $1.87938$ & $-$  \\
$m_4=2m_1\cos(\pi/18)$ & $1.96961$ & $+$   \\
$m_5=4m_1\cos(\pi/18)\cos(5\pi/18)$ & $2.53208$ & $+$  \\
$m_6=4m_1\cos(4\pi/18)\cos(2\pi/18)$\qquad\ & $2.87938$ \qquad\qquad &
$-$  \\
$m_7=4m_1\cos(\pi/18)\cos(2\pi/18)$ & $3.70166$ & $+$   \\
\hline
\end{tabular}
\end{indented}
\end{table}

Note that $A_1$, $A_3$ and $A_6$ are odd under spin reversal symmetry while
the other particles are even. Only the first four particles have a mass below
the stability thresholds $2 m_1$ and $m_1+m_2$. The higher three particles are nevertheless
stable along this axis due to the integrability
of the theory.

\subsection{Decay processes}
\begin{figure}[t!]
\begin{center}
\begin{tabular}{cc}
\includegraphics[scale=0.6]{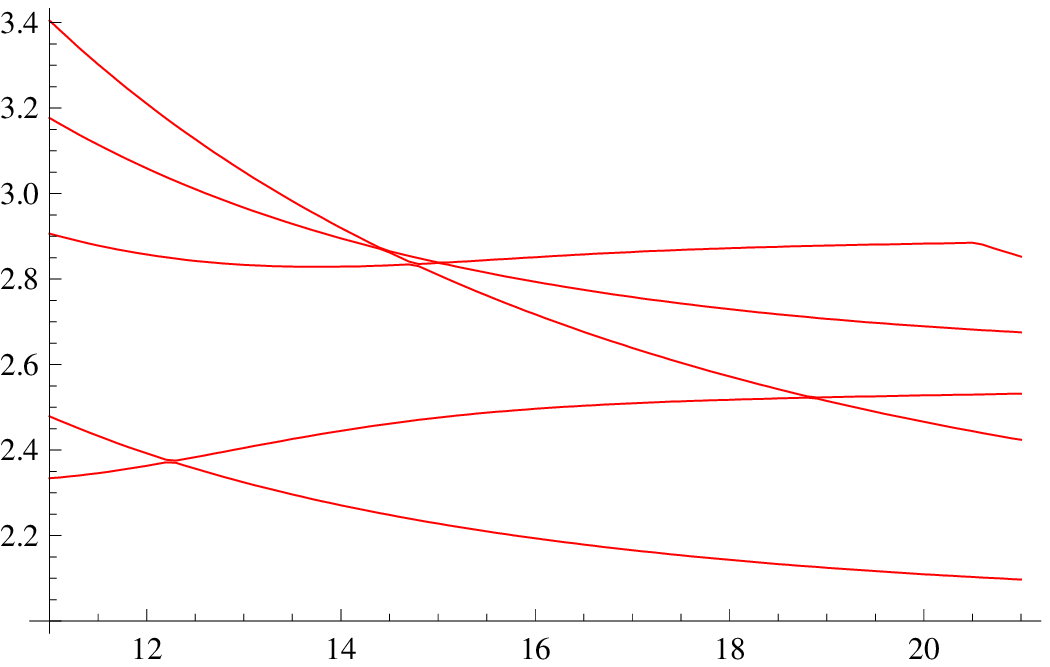}&
\includegraphics[scale=0.6]{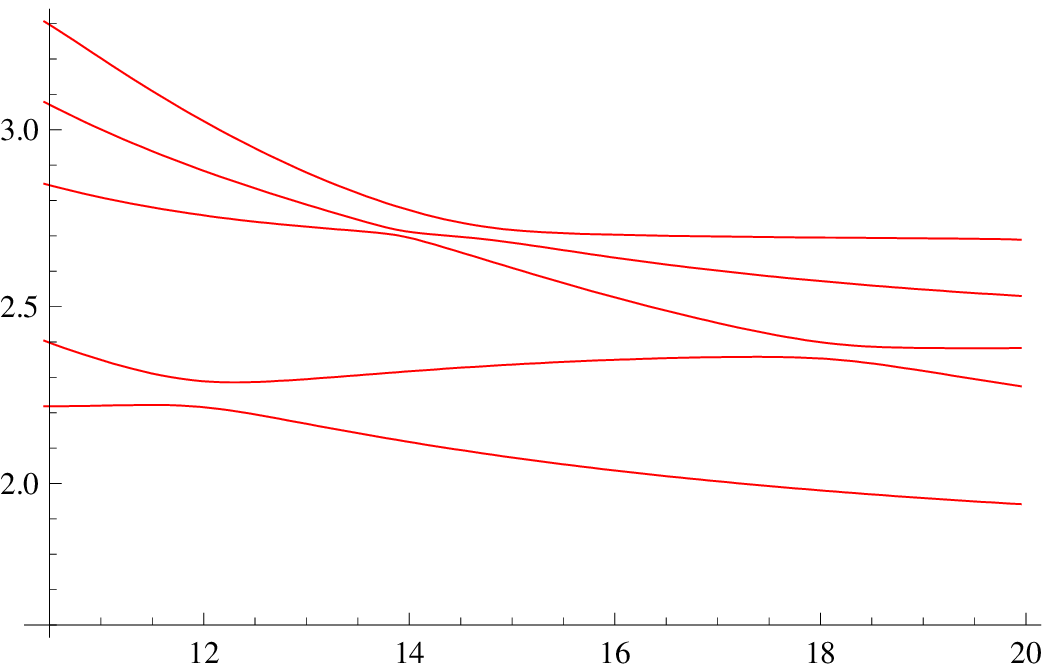}\\
(a) & (b)
\end{tabular}
\caption{
\label{decay}
{\em Signature of the appearance of a resonance out of the integrability in the finite volume spectrum.
(a) Integrable point. (b) Perturbation switched on. These are the plots of the energy levels versus $mR$, where $m$ is the mass gap. }}
\end{center}
\end{figure}

Moving away from the axis $\eta_+ =0$, the expectation is that there exists a range of values of
$\eta_+$ where the four lowest particles are still stable, whereas the higher three particles decay in all possible channels compatible with the symmetry of the perturbation. The perturbation that breaks the integrability of the theory at $\eta_+=0$ and plays the role of the operator $\Upsilon(x)$ in eqn
(\ref{perturbedaction}) is the field $t$, which is even under the $\ZZ_2$ spin reversal symmetry. This implies certain selection rules in the decay processes. So, for instance, the particle $A_5$ can only decay in the channel
\begin{equation}
\label{eq.dec1}
A_5 \rightarrow A_1 \,A_1 \, ,
\end{equation}
even though the decay process $A_5 \rightarrow A_1\,A_2$ would be permitted by
kinematics.
Similarly, it is easy to check that the other possible decay processes compatible with the selection rule are
\begin{eqnarray}
&& A_6 \rightarrow A_1\, A_2 \nonumber \\
&& A_7 \rightarrow A_1 \,A_1 \nonumber \\
&& A_7 \rightarrow A_2 \, A_2 \label{eq.dec2} \\
&& A_7 \rightarrow A_1 \, A_3 \nonumber \\
&& A_7 \rightarrow A_2 \, A_4 \nonumber \\
&& A_7 \rightarrow A_1\,A_1\,A_2\, . \nonumber
\end{eqnarray}
The decay processes have a fingerprint in the finite volume
spectra, namely the repulsion of the energy lines \cite{DMS}. Observe, for instance, Figure \ref{decay}.a: this refers to the lowest energy difference lines above the two-particle threshold (in the even sector only) on the integrable $E_7$ axis. One can see that there are several level crossings; in particular
there is one around $mR \sim 12$ between the first and the second
lines. The first line corresponds to the threshold state $\ket{A_1
A_1}$, whereas the second line corresponds to the particle
$\ket{A_5}$. 
When we switch on the $g_4$ perturbation,
the crossing disappears, as it can clearly be seen  in Figure
\ref{decay}.b. In particular, the repulsion of the first and
second lines at $mR \sim 12$ corresponds to the decay $A_5 \to
A_1A_1$. 

In principle, the widths $\Gamma^k_{ij}$ of the decays $A_k \to A_iA_j$ can be
 determined from the repulsion of energy lines
 by a method proposed in \cite{PT}.
In this approach, which was called Breit-Wigner method in \cite{PT}, $\Gamma$
 is related to the two-particle phase shifts  
of the various two-particle levels $\ket{A_iA_j} $. In more detail, in order to measure $\Gamma$ one can use the formula
\begin{equation} 
\label{shift}
\Delta \delta= {\rm min}\, \delta_1(E) - {\rm max} \,
\delta_{2}(E)=4 \sqrt{-\beta \Gamma}\ ,
\end{equation}
where 
\begin{equation} \beta= \left.  \frac{{\rm d} \delta_0}{{\rm d}E
}\right| _{E=E_k}  
\end{equation} 
should be negative, $E_k$ is the center of the resonance and $\delta _0(E)$ is defined by
the integrable S-matrix as 
\beq 
\delta_{0}(E) = - \rmi {\rm \, ln\, }
S(E)\ .
\eeq 
For the sake of clarity, we have suppressed the labels $i$, $j$, $k$ relative
to the particles. 
The symbols  $\delta_1(E)$,   $\delta_2(E)$ in (\ref{shift}) refer to the
phase shifts of
two neighbouring  two-particle 
levels $\ket{A_iA_j}$ and are given by
\beq 
\delta_a(E)\,=\, - R(E)\, p_a(E), \qquad a=1,2\ ,  
\label{important}
\eeq
where $R$ is the circumference of the cylinder and $p_a(E)$ is the momentum,
defined by the equation 
\begin{equation}
E \,=\,\sqrt{p_a^2+m_i^2}+\sqrt{p_a^2+m_j^2}\ .
\end{equation}
Hence, from the numerical determination of the energy levels we can measure
$E(R)$ as a function of $R$; 
from this measurement, using 
(\ref{important}), we can extract $\delta_1(E)$ and $\delta_{2}(E)$,
 and then (\ref{shift}) provides  $\Gamma$.  

Using the above consideration we have studied the decay processes 
\begin{eqnarray*}
&& A_5 \to A_1A_1\\
&& A_6 \to A_1A_2 
\end{eqnarray*}
at $g_2=1$ and small $g_4$, finding the following numerical values for the decay rates 
\beq
\Gamma^5_{11}\sim 2\cdot g_4^2\ ,\qquad \Gamma^6_{12}\sim 0.2\cdot g_4^2\ , 
\eeq
and therefore the universal ratio 
\beq 
\frac{\Gamma^6_{12}}{\Gamma^5_{11}} 
\,\sim \, 0.1\ .
\label{urdecay}
\eeq
This result shows that the decay of $A_6$ is considerably slower than the decay of $A_5$. Although the kinematical phase space factor $\Phi^M_{m_i,m_j}$  generally tends to suppress the decay of heavier particles in $1+1$ dimensions \cite{DGM}, in the present case this effect is not sufficient, however, to justify the result above: in fact, the phase space estimate of the universal ratio (\ref{urdecay}) gives in this case 
\beq
\frac{\Gamma^6_{12}}{\Gamma^5_{11}} 
\sim \frac{\Phi^6_{12}}{\Phi^5_{11}} \,=\, 
\sqrt{\frac{m_5^2 (m_5^2 - 4 m_1^2)}{(m_6^2 - (m_1-m_2)^2) (m_6^2 - (m_1 +
    m_2)^2)}} =   0.783...\ .
\eeq 
Therefore the longer lifetime of $A_6$ has to be attributed to the dynamics of the model. We plan to analyze this aspect in a future publication. 

\subsection{Mass corrections}

Coming back to the stable particles of the spectrum,
it is relatively easy to compute the correction to the masses of the two
lowest particles once we move away from the $E_7$ integrable axis. This can be
done by combining the FFPT and the TCSA, together with a 
cross-checking between the two
methods. The first order mass correction of the particle $A_i$  moving away from the
$\eta_+ = 0$ axis is given by
\begin{equation}
\label{eq.ffpt1}
\delta m_i=\frac{g_4}{m_i}2\pi\lambda_4 F^{t}_{ii}(\rmi\pi),
\end{equation}
where $F^{t}_{ii}(\theta_1-\theta_2)=\brakettt{0}{t(0)}{A_i(\theta_1)
  A_i(\theta_2)}$ is the two-particle
form factor of $t$. As a matter of fact,
$F^{t}_{11}(\rmi\pi)$, $F^{t}_{22}(\rmi\pi)$,  $F^{t}_{33}(\rmi\pi)$ and  $F^{t}_{44}(\rmi\pi)$ can be directly extracted by TCSA
(see Appendix B and \cite{GuiMagn,FMS}). At $g_2=1$ we obtained
\begin{eqnarray}
\label{eq.kk}
F^{t}_{11}(\rmi\pi) & =   - \frac{1.28}{2\pi\lambda_4}\ ,\qquad
F^{t}_{22}(\rmi\pi) & =  - \frac{1.25}{2\pi\lambda_4}\ ,\\
\label{eq.kk2}
F^{t}_{33}(\rmi\pi) & =  - \frac{2.03}{2\pi\lambda_4}\ ,\qquad
F^{t}_{44}(\rmi\pi) & =  - \frac{3.41}{2\pi\lambda_4}\ .
\end{eqnarray}
Similarly, from TCSA we can also compute the absolute value of the one-particle form factors $F_2^{t} = \brakettt{0}{t(0)}{A_2(\theta)}$ and $F_4^{t} = \brakettt{0}{t(0)}{A_4(\theta)}$. The result is
\beq
\label{eqf.f2}
|F_2^{t}|\,=\,\frac{0.35}{2\pi\lambda_4}\ ,\qquad
|F_4^{t}|\,=\,\frac{0.26}{2\pi\lambda_4}\ .
\eeq
As shown in the Appendix A, there is an exact relation between the two-particle form factors $F^t_{11}(i\pi)$ and $F^t_{22}(i\pi)$ and the one-particle form factors $F^t_2$ and $F^t_4$, expressed by
\begin{eqnarray}
\label{eq.aa}
F_{11}^{t}(\rmi\pi) & = 11.397\cdot F_2^{t}
 + 10.3632\cdot F_4^{t}\\
\label{eq.bb}
F_{22}^{t}(\rmi\pi) & =  23.1056\cdot F_2^{t}
 + 26.2404\cdot F_4^{t}\ .
\end{eqnarray}
If we assume that $F_2^{t} < 0$ and  $F_4^{t} > 0$,
then (\ref{eqf.f2}),  (\ref{eq.aa}) and (\ref{eq.bb})
yield
\begin{equation}
F^{t}_{11}(\rmi\pi)\,=\,\frac{-1.29}{2\pi\lambda_4}\ ,\qquad
F^{t}_{22}(\rmi\pi)\,=\,\frac{-1.26}{2\pi\lambda_4}\ ,
\end{equation}
which is in satisfactory agreement with (\ref{eq.kk}).

With the above values, we can extract the universal ratios
\begin{eqnarray}
&& \delta m_2/\delta m_1 = 0.76\ , \nonumber \\
&& \delta m_3/\delta m_1 = 0.84\ , \\
&& \delta m_4/\delta m_1 = 1.35\ . \nonumber 
\end{eqnarray}
In order to get another universal ratio, let us look at the first order
correction to the vacuum energy density $\epsilon_{vac}$. By TCSA (see
Appendix B) we obtain
\beq
\label{eq.vd}
\frac{\delta \epsilon_{vac}}{g_4}=0.11\ .
\eeq
On the other hand, the exact value of this quantity is
\beq
\frac{\delta \epsilon_{vac}}{g_4}=0.1130... =
\Lambda_4\times
(2\pi)^{\frac{1-\Delta_2-\Delta_4}{1-\Delta_2}}
\lambda_2^{\frac{\Delta_4}{1-\Delta_2}} \lambda_4\ ,
\eeq
which is in good agreement with the above numerical result.
$\Lambda_4=3.70708...$ is the vacuum expectation value of $t$ calculated in
\cite{FLZZ,FMS} (the factor next to $\Lambda_4$ arises from various
normalization factors
used in the present paper and  in \cite{FLZZ,FMS}).
This allows us also to extract the universal ratio
\begin{equation}
\frac{\delta \epsilon_{vac}}{m_1\delta m_1} = - 0.086\ .
\end{equation}

\subsection{The first quadrant}
\label{firstquadrant}

After the analysis in the previous section we can now proceed to investigate the evolution of the spectrum in the first quadrant. First of all,
note that the corrections to the two lowest masses have the same sign, which depends on the sign of $g_4$. We can use the above results for $\delta m_1$ and $\delta m_2$ to estimate the value of
$\eta_+^{(2)}$ where the mass $m_2(\eta_+)$ of the second particle reaches the threshold $2 m_1(\eta_+)$ of the lowest particle. This happens for a value of
$\eta_+ > 0$ of the first quadrant, given by
$\eta_+^{(2)} \simeq 1$: although this is higher than the value $\eta_+^{(2)} \simeq 0.7$ extracted from the numerical data shown in Figure \ref{fig.ratios}, it is nevertheless a reasonable estimation of this quantity since the theoretical result was based on just first-order perturbation theory.

Besides the three higher particles $A_5, A_6,A_7$ that were stable only at
$\eta_+=0$ and decay as soon as we move away from the horizontal axis, the
TCSA analysis shows that the same pattern as was seen in the fourth quadrant also
persists in the first quadrant: namely, increasing the value of $\eta_+$, the
number of stable particles continues to decrease. The first particle that
disappears into the threshold $2 m_1(\eta_+)$ is $A_4$ at the critical
value $\eta_+^{(4)}$, which can be estimated from first-order perturbation
theory to be $\eta_+^{(4)}\simeq 0.04$. This is followed by $A_3$ that disappears into its
lowest threshold, given by ($m_1(\eta_+) + m_2 (\eta_+))$, at the
critical value
$\eta_+^{(3)}$, for which perturbation theory gives $\eta_+^{(3)}\simeq
0.62$. These theoretical results for $\eta_+^{(3)}$ and $\eta_+^{(4)}$  are in
reasonably good agreement with the
numerical data (Figure  \ref{fig.ratios}). An example of the finite volume
spectra calculated numerically is shown in Figure \ref{h2}.a
 at $\eta_+=0.3$, where
the theory contains $3$ stable particles.

As just described, moving toward the positive vertical axis, there is a depletion of the number of stable particles, until there remains only one in the neighbourhood of $\eta_+ = + \infty$. At the same
time, notice that increasing $\eta_+$, the value of this lowest mass becomes also smaller\footnote[4]{This behavior is in agreement
with the first-order
  correction (\ref{eq.ffpt1}) to $m_1$, that is negative for $g_4 > 0$.}.
 The mass gap of the theory finally vanishes when $\eta_+ = + \infty$, i.e.\
 when we have reached the vertical axis.

During this evolution, however, there has also been a qualitative change of
the spectrum: in fact, reaching the positive vertical axis the lowest 
excitation has turned into a fermion. 
This becomes evident by looking directly at the nature of the theory at
$\eta_+ = + \infty$ 
(notice that the positive vertical axis also corresponds to $\eta_- =  +\infty$).

\subsection{The spontaneously SUSY breaking axis}

The analytic control of the theory in the vicinity of $\eta_{\pm} = + \infty$
 is provided by the integrability of the model along the vertical axis, where
 the system is also supersymmetric.
 Along the
positive vertical axis the supersymmetry is however  spontaneously broken: the
 low-energy excitations are given in this case by the massless right and left
 mover Majorana fermions, which play the role of goldstino
 \cite{ZZ,Z2,KMS}. The factorized scattering theory was proposed in \cite{Z2}
 and the basic form factors were calculated in \cite{DMS2}. The massless Majorana
 fermions are nothing else than those of the Ising model,
 connected to the TIM by the massless renormalization group flow that occurs along this line.

Breaking the integrability of the $\eta_{\pm} = \infty$ model by means of the
operator $\Upsilon =\varepsilon$, the left and right moving excitations become
adiabatically massive, as it can be directly observed in the TCSA data (see
Figure \ref{fig.absmass}).
To compute the mass $m$ of the fermion generated by the perturbation, we need to employ in this case the massless FFPT \cite{CM}: at the lowest order, we have
\begin{equation}
\label{deltammassless}
 m
\simeq g_2 \lim_{\theta_{RL}\to- \infty} F_{RL}^{\varepsilon}(\rmi\pi
+\theta_{RL})\ ,
\end{equation}
where the two-particle (right-left) form factor of the $\varepsilon$ operator is given by \cite{DMS2}
\begin{equation}
F^\varepsilon_{RL}(\theta) \,=\,
\exp \left ( \frac{\theta}{4}-\int_0^{\infty} \frac{\intd t}{t}
\frac{\sin^2 \left ( \frac{(\rmi\pi - \theta)t}{2 \pi} \right )}
{\sinh t \cosh \frac{t}{2} } \right )\ .
\label{epsilon}
\end{equation}
Using (\ref{deltammassless}) and (\ref{epsilon}),
one can easily check that $m$ is a {\em finite} quantity, positive for $g_2 >0 $ and negative for $g_2 <0$.
Note that, for a Majorana fermion, a negative value of the mass signals that we are in a low temperature phase, i.e.\ that the theory has two degenerate vacua. This is the topic of the next section.

\section{The spectrum in the low temperature phase}

The low temperature phase is composed of the second and the third
quadrants. This phase is described by the variable $\eta_- \in
(-\infty,+\infty)$: $\eta_- = \infty$ corresponds 
to the positive vertical axis, $\eta_-=0$ to the negative horizontal axis and $\eta_-=-\infty$ to the negative vertical axis.

In this phase the model presents generically two degenerate ground states
 $\ket{-}$ and $\ket{+}$, which are mapped into each other by the
 spin reversal symmetry operation $Q$: $\ket{+} = Q \ket{-}$. An important feature that
 is worth stressing is the following: no matter how we vary the coupling constants
 in this parameter region, the two vacua $\ket{-}$ and $\ket{+}$ always remain 
 degenerate, the only changes being in the shape and the height of the barrier
 between them. Hence, in this model the kink states that interpolate between
 the two vacua are always stable excitations.  As we will comment below, this
 can be explicitly 
checked and confirmed by a FFPT computation at $\eta_- =0$, where the theory has the $E_7$ structure.

 The only points where one has to be careful are the two limiting cases
 $\eta_- =  \infty$ and $\eta_-=-\infty$. In the former case, the barrier
 between the two vacua disappears and, as we saw at the end of Section
 \ref{firstquadrant}, the kinks become massless. In the latter case, 
the two vacua $\ket{-}$ and $\ket{+}$ and 
 the false vacuum $\ket{0}$ that emerges between them
 in the region $\eta_- < 0$ become degenerate precisely at $\eta_-
 = -\infty$. Hence, at this point, the original kinks break into a new set of
 smaller kinks, i.e.\ those relative
 to the three vacua of the first order 
phase transition of the TIM.

To obtain the spectrum of the theory in the low temperature phase we can take
advantage of the duality of the model. At a formal level, note that the masses
at a given point $(g_2,g_4)$ in the low temperature phase are the same as the
masses in the dual point $(-g_2,g_4)$ in the high temperature phase. This
result can be obtained as follows: since $\varepsilon$ is odd and $t$ is 
even under the duality transformation $D$: 
\begin{eqnarray}
 D^{-1}\varepsilon D & = -\varepsilon, \nonumber \\
 D^{-1} t D & =  t,\nonumber
\end{eqnarray}
for the Hamiltonian we have
\beq
D^{-1}H(g_2,g_4) D \,=\, H(-g_2,g_4)\ .
\eeq
However, to interpret correctly the TCSA data, one needs to take into account the existence of kink states and the periodic boundary conditions imposed to the system.

\begin{figure}
\begin{center}
\begin{tabular}{c}
\includegraphics[
  scale=0.85,
  angle=0]{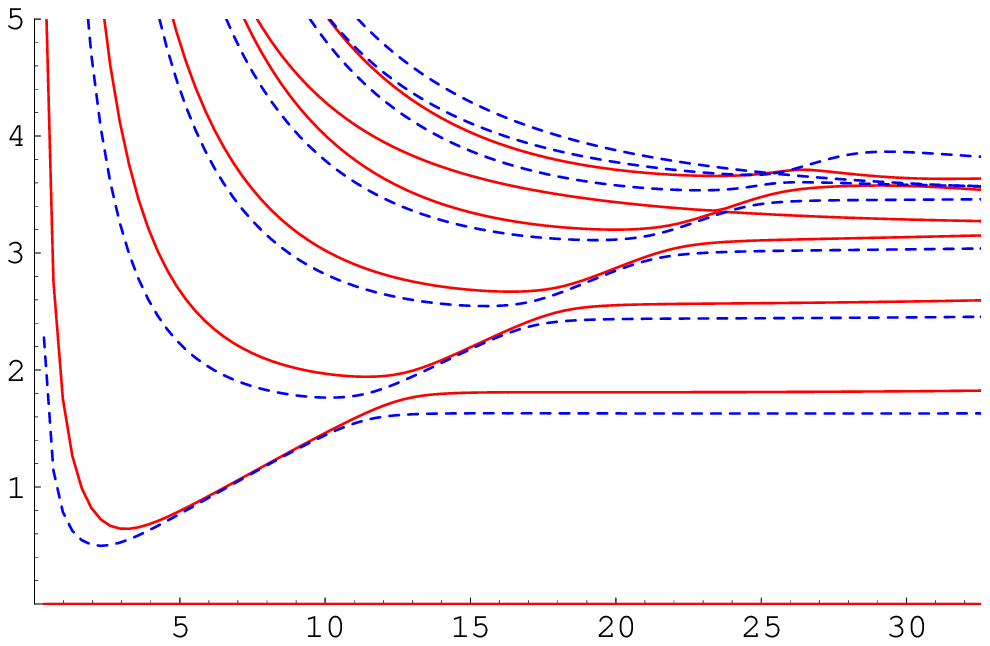}\\
(a) $g_2=1$, $g_4=-0.4$\\[5mm]
\includegraphics[
  scale=0.85,
  angle=0]{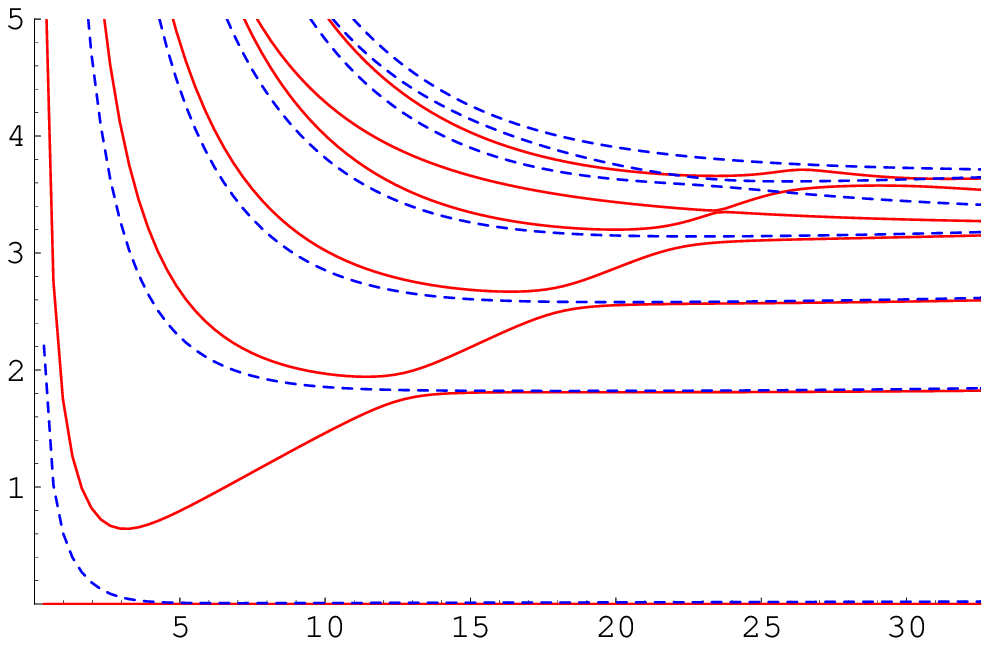}\\
(b)
$g_2=-1$, $g_4=-0.4$
\end{tabular}
\vspace{3mm}
\caption{\label{h1}
{\em
The first $14$ energy level differences $E_i-E_0$, $i=0...13$, as functions of $mR$ in the (a) 4th and (b) 3rd quadrants at the dual values of the couplings constants. Even levels are in red, odd levels are in blue with dashed line. $m$ denotes the mass gap.}}
\end{center}
\end{figure}

\begin{figure}
\begin{center}
\begin{tabular}{c}
\includegraphics[
  scale=0.85,
  angle=0]{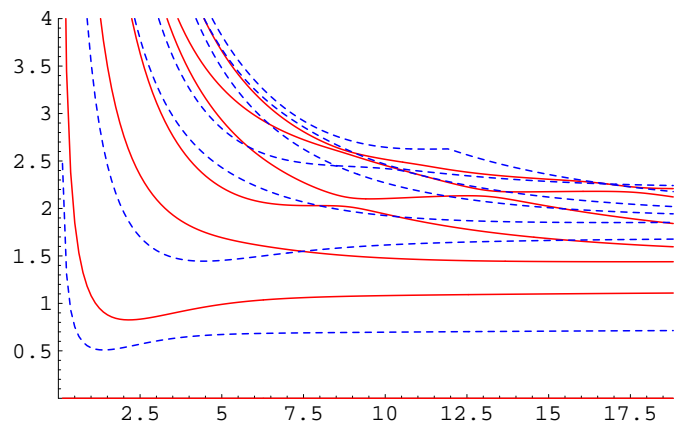}\\
(a) $g_2=1$, $g_4=0.3$\\[5mm]
\includegraphics[
  scale=0.85,
  angle=0]{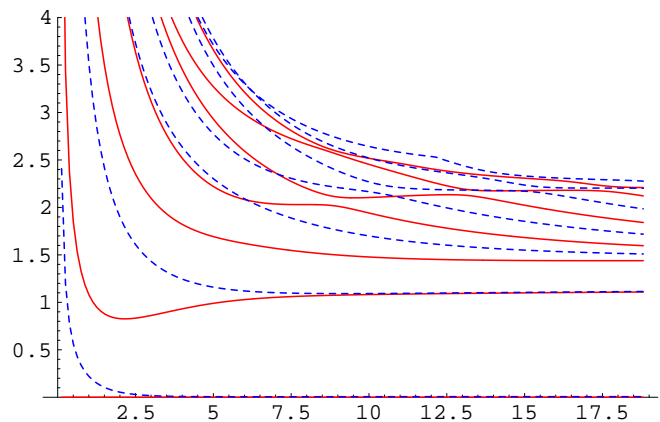}\\
(b)
$g_2=-1$, $g_4=0.3$
\end{tabular}
\vspace{3mm}
\caption{\label{h2}
{\em The first $14$ energy level differences $E_i-E_0$, $i=0...13$, as
  functions of $mR$ in the (a) 1st and (b) 2nd quadrants at the dual values of
  the coupling constants. Even levels are in red, odd levels are in blue with dashed line. $m$ denotes the mass gap.}}
\end{center}
\end{figure}

Note that $D$ maps the even sector of the Hilbert space onto itself, so in the
even sector $H(g_2,g_4)$ has the same spectrum as $ H(-g_2,g_4)$. Concerning
the odd sector of the spectrum, the particles with odd parity in the high
temperature phase 
become instead kinks in the low temperature
 phase, i.e.\ topologically charged states that interpolate between the two
 vacua. However, the periodic boundary conditions that we have imposed on the
 TCSA filter only topologically neutral states. Hence, by switching $g_2
 \rightarrow - g_2$, we expect that the net result in the numerical outcomes
 will be the disappearing of the energy levels corresponding to the odd
 particles in the high temperature phase: in the low temperature phase these
 lines should become exponentially degenerate with the energy lines of the
 even levels. In other words, in the low temperature 
phase the TCSA data
should show doubly degenerate lines and each doublet should have an even and
odd member. This is indeed the case, as clearly shown in Figures \ref{h1} and
\ref{h2}, where we present the spectrum of the theory at given values
$(g_2,g_4)$ and $(-g_2,g_4)$. This feature is nothing but 
a finite volume manifestation of the spontaneous breaking of the $\ZZ_2$ spin reversal symmetry: the two vacua (and all the excitations above them) have in fact an exponential splitting of their energy for the tunneling phenomenon that occurs at finite volume.

Summarizing the considerations above, in the low temperature phase and at
infinite volume the model has two ground states $\ket{+}$, $\ket{-}$, each
ground state has a sequence $\ket{C_n}_{\pm}$ of topologically neutral
particles above it, and there is also a sequence of kinks and antikinks
$\ket{\mathcal{D}^{m}_{+-}}$,  $\ket{\mathcal{D}^{m}_{-+}}$ which interpolate
between the vacua\footnote[5]{The indices $n$ and $m$ number the members of the
  sequences, while $+$ and $-$ refer to the vacua between which the kinks
  interpolate or which support the neutral states.}. Under 
spin reversal symmetry the ground states, particles and kinks have the following property: 
\begin{eqnarray*}
&& \ket{\pm } \,=\, Q \ket{\mp}\ ; \\
&& \ket{C_{n}}_{\pm} \,=\, Q \ket{C_{n}}_{\mp}\ ; \\
&& \ket{\mathcal{D}^{n}_{+-}} \,=\, Q \, \ket{\mathcal{D}^{n}_{-+}}\ . 
\end{eqnarray*}
The masses of the neutral states at a given point of the plane of the coupling
constants are the same as the masses of the even particles in the high 
temperature phase at the dual point, whereas the masses 
of the kinks are the same as the masses of the odd particles in the high temperature phase.
In finite volume with periodic boundary conditions, however, we do not see
individual kinks or other topologically charged configurations.  We do see the
vacua, the neutral particles and only neutral configurations of two or other
even number of free kinks, which, moreover, get mixed in finite volume into
even and odd eigenstates of the spin reversal operator and acquire an
exponential splitting due to tunneling effects. With periodic boundary
conditions there are also identifications between certain configurations,
similarly as in the first order transition point (see Figure \ref{3vacua}).

After  the  discussion above we can  proceed to describe the spectrum in the
second and third quadrants in detail.

\subsection{The third quadrant}

At $\eta_-=-\infty$ the model is at the first order phase transition point, which was described in Section \ref{sec.fourth}. As we move into the third
quadrant by switching on the perturbation $g_2 \int \varepsilon (x) \intd x$ $(g_2 < 0)$, the degeneracy of the three vacua is lifted. In contrast with the fourth quadrant, the central vacuum $\ket{0}$ now becomes a metastable ground state, and the two other vacua $\ket{\pm}$ remain true ground states. The gap $\Delta E$ separating  $\ket{0}$ from  $\ket{\pm}$ at a given point is equal to the gap at the dual point in the fourth quadrant. The expected consequence of the lifting of the degeneracy of the vacua is again the confinement of the kinks of the unperturbed system, which is confirmed by the
same FFPT argument as for the fourth quadrant. The linear confining potential
between the kinks gives rise to a dense sequence of bound states. As in the
fourth quadrant, the number of these bound states goes to infinity as $g_2\to 0$. In the third quadrant, however, the two-kink states $\ket{K_{0-}K_{-0}}$ and
$\ket{K_{0+}K_{+0}}$ are those which disappear from the spectrum, and the bound states arise from the other two-kink states
 $\ket{K_{-0}K_{0+}}$,
$\ket{K_{+0}K_{0-}}$,
 $\ket{K_{-0}K_{0-}}$ and
$\ket{K_{+0}K_{0+}}$. Some of the bound states above are topologically charged, i.e.\ they are kinks,  while others are neutral. The neutral particles arise from the neutral two-kink configurations:
\beq
\label{eq.c1}
\ket{C^{n}}_+ \sim \ket{K_{+0}K_{0+}},\qquad
\ket{C^{n}}_- \sim \ket{K_{-0}K_{0-}}.
\eeq
It is important to note that $\ket{C^{n}}_+$ and 
$\ket{C^{n}}_-$ are degenerate for any given
$n$, due to the spontaneously broken $\ZZ_2$ symmetry, although  their degeneracy is exponentially lifted in finite volume. The topologically  charged particles arise from topologically charged two-kink states:
\beq
\label{eq.d1}
\ket{\mathcal{D}^n_{-+}} \sim \ket{K_{-0} K_{0+}} ,\qquad \ket{\mathcal{D}^n_{-+}} \sim \ket{K_{-0} K_{0+}},
\eeq
where the subscripts $-+$ and $+-$ in $\ket{\mathcal{D}^n_{-+}}$ and $\ket{\mathcal{D}^n_{+-}}$
denote the vacua between which these kink configurations  interpolate.
 $\ket{\mathcal{D}^n_{-+}}$ and $\ket{\mathcal{D}^n_{+-}}$ are mapped into
each other by the 
spin reversal: $Q\ket{\mathcal{D}^n_{-+}} = \ket{\mathcal{D}^n_{+-}}$, therefore they have equal mass.

As we mentioned, under duality an even particle $\ket{B_n^+}$ in the high temperature phase
corresponds to the neutral particles $\ket{C^n}_{\pm}$ of the same mass in the low temperature phase, and an odd particle  $\ket{B_n^-}$ corresponds to a
kink-antikink pair $\ket{\mathcal{D}^n_{+-}}$,  $\ket{\mathcal{D}^n_{-+}}$ of the same mass.

The neutral particles $\ket{C^n}_{\pm}$ (more precisely, their even and odd superpositions) can indeed be seen in the TCSA spectra, whereas the kinks  $\ket{\mathcal{D}^n_{+-}}$ and  $\ket{\mathcal{D}^n_{-+}}$ are filtered out by the
periodic boundary conditions. The neutral two-kink states constituted by  $\ket{\mathcal{D}^n_{+-}}$ and  $\ket{\mathcal{D}^n_{-+}}$ can nevertheless be observed in the numerical data. The disappearance of the original two-kink states  $\ket{K_{0-}K_{-0}}$ and $\ket{K_{0+}K_{+0}}$ does not have drastic effect on the spectra with periodic boundary conditions for similar reasons as for the third quadrant.

The  false vacuum with even parity can also be seen  in the  $\eta_- < 0$
domain  in the TCSA spectra in the form of a  linearly rising line-like
pattern (see Figure \ref{h1}.b). The duality relates this metastable vacuum to the metastable vacua in the first quadrant.

\subsection{The $E_7$-related line}
\label{sec.6x}

The number of stable particles at any given point
in the low temperature phase is equal to the number of particles in the dual point in the high temperature phase, and this number decreases as
$\eta_-$ is increased. The number of stable particles around $\eta_-=0$ is $4$, however at $\eta_-=0$, i.e.\ on the negative horizontal axis, $3$ more particles become stable above threshold. These particles are related by duality to the three highest particles existing at  $\eta_+=0$. In particular, there
are two neutral particles $\ket{C^5}_{\pm}$, 
$\ket{C^7}_{\pm}$ of masses $m_5$ and $m_7$ (see Table \ref{tab.E7}) which correspond to the even particles $\ket{A_5}$, $\ket{A_7}$, and there
is one kink-antikink pair $\ket{\mathcal{D}^6_{+-}}$,
 $\ket {\mathcal{D}^n_{-+}}$ of mass $m_6$
corresponding to the odd particle $\ket{A_6}$.

It was stressed in the introduction that an important property of the model is
the presence of kinks in the low temperature phase which do not get confined
under perturbations. At $\eta_- = 0$ this non-confinement can be  explicitly
verified by FFPT. For this, one has to show that the form
factors giving the first-order corrections to the masses of
$\ket{\mathcal{D}^i_{+-}}$ and  $\ket{\mathcal{D}^i_{-+}}$ are finite.  This
is indeed the case, 
since these form factors  
\begin{eqnarray*}
&& \brakettt{-} {t(0)} {\mathcal{D}^i_{-+}(\rmi\pi) \,
\mathcal{D}^i_{+-}(0)}\ , \\
&& \brakettt{+} {t(0)} {\mathcal{D}^i_{+-}(\rmi\pi) \,
\mathcal{D}^i_{-+}(0)}\ , 
\end{eqnarray*}
(where $i =1,3$) are equal by duality to the form factors
 $\brakettt{0}{t(0)}{A_{i}(\rmi\pi)A_{i}(0)}$, and the latter are finite since
 $t$ does not have non-trivial
 semi-local index with respect to the $\ket{A_i}$ particles, which are local excitations. 

The particles  $\ket{C^5}_{\pm}$,
$\ket{C^7}_{\pm}$,  $\ket{\mathcal{D}^6_{+-}}$,
 $\ket{\mathcal{D}^6_{-+}}$ are unstable near the $\eta_-=0$ axis, the allowed decay processes
can be obtained from (\ref{eq.dec1}) and (\ref{eq.dec2}) by replacing $A_i$ by the
appropriate dual particles. For example, (\ref{eq.dec1}) corresponds to
\beq
C^5_{+} \,\to\, \mathcal{D}^1_{+-} \mathcal{D}^1_{-+}\ ,\qquad
C^5_{-} \,\to\, \mathcal{D}^1_{-+} \mathcal{D}^1_{+-}\ .
\eeq

\subsection{The second quadrant}

Similarly to the third quadrant, the particle spectrum in the second quadrant
is related by duality to the spectrum in the first quadrant. Even particles
in the first quadrant correspond to neutral particles, odd particles
correspond to kinks.
The number of particles decreases from $4$ to $1$ as $\eta_-$ increases from
$0$ to $\infty$, the critical values where the particles and kinks cross the threshold
are the same as in the first quadrant, i.e.\ $\eta_+^{(4)}$, $\eta_+^{(3)}$,
$\eta_+^{(2)}$. For $\eta_-> \eta_+^{(2)}$ there is only one kink-antikink
pair in
the spectrum. Similarly to the first quadrant,
metastable vacua are not present.
 As $\eta_-$ goes to infinity, the potential barrier between the two vacua
$\ket{\pm}$ and the mass of the kink-antikink pair decrease until they finally
 vanish at the second-order transition point $\eta_-=\infty$.
An example of the finite volume spectra calculated numerically is shown in
Figure \ref{h2}.b at $\eta_-=0.3$, which is dual to the point $\eta_+=0.3$
that the example Figure \ref{h2}.a for the first quadrant shows.

\section{Conclusions}

In this paper we have studied the particle spectrum of the TIM with spin
 reversal symmetric perturbations. This is the simplest bosonic non-integrable
 field theory where kink excitations do not get confined by changing the
 coupling constants
 over a wide range of values. By varying the couplings, the model interpolates
 between a SUSY theory (either in its exact or in its spontaneouly broken phase) and
 a theory ruled by the exceptional algebra $E_7$, with a spectrum given by
 purely scalar particles (in the high temperature phase) and kinks and bound
 states thereof (in the low temperature phase).  

Our results, obtained both by the FFPT and TCSA methods, are also in agreement
 with the Landau-Ginzburg picture of the TIM. We have found that the particle
 masses are the same at dual points in the high and low temperature phases,
 whereas the spin reversal symmetry properties and the topological charges of
 the particles are different in the two phases: even particles in the high temperature phase correspond to topologically neutral particles in the low temperature phase, odd particles correspond to (topologically charged) kink states. The number of
stable particles tends to infinity in the vicinity of the first-order
transition line, which is similar to the accumulation of particles found in
the low temperature Ising model perturbed by a magnetic field \cite{McWu}.
However, in contrast with the Ising model, in TIM the parity is a good quantum
number and the corresponding particles emerging from the kink-antikink
threshold
 carry  even or odd parity eigenvalues. 
  
We have also computed the first-order corrections
to the lowest four masses 
and the vacuum energy density at the $E_7$-related line, as well as the
corrections to the energy densities of the three degenerate vacua 
at the first-order phase transition line. 
All these analytic results are in good agreement with 
the numerical estimates extracted by means of the TCSA. 

Finally, we would like to mention that the three-frequency
sine-Gordon model also  has  a tricritical point, the neighbourhood of which
was studied by the TCSA in \cite{Toth,Toth2}. In the light of the results of
this paper, it would be interesting to investigate more thoroughly the first
order transition in the three-frequency sine-Gordon model and, in particular,
to study the evolution of the particle spectrum of this model. Moreover, the
strategy adopted in the present paper can be also used to study the spectrum of higher multi-critical theories, described by higher minimal models of conformal field theories perturbed by several operators.

\section*{Acknowledgements}

We would like to thank Gesualdo Delfino and Robert Konik for useful discussions.

\newpage

\appendix

\appsection
{\bf Form Factor Bootstrap}

\noindent
In this appendix we explain the derivation of the relations (\ref{eq.aa}) and (\ref{eq.bb}) by
the form factor approach. The two-particle form factor $F^{t}_{ii}(\theta)$ can be written as \cite{DM}
\begin{equation}
F_{ii}^{t}(\theta)=Q_{ii}^{t}(\theta)\frac{F_{ii}^{min}(\theta)}{D_{ii}(\theta)},
\end{equation}
where $F_{ii}^{min}(\theta)$ and  $D_{ii}(\theta)$ can be obtained by specializing the general formulas \cite{DM} to the present case:
\begin{eqnarray}
D_{11}(\theta) &=P_{10}(\theta)P_{2}(\theta) \nonumber \\
D_{22}(\theta) &=P_{12}(\theta)P_{8}(\theta)P_{2}(\theta)\\
F_{11}^{min}(\theta) &=-\rmi\sinh\frac{\theta}{2}\, g_{10}(\theta)g_2(\theta) \nonumber \\
F_{22}^{min}(\theta) &=-\rmi\sinh\frac{\theta}{2}\,
g_{12}(\theta)g_8(\theta)g_2(\theta)\ ,\nonumber 
\end{eqnarray}
where
\beq
P_{n}(\theta)\,=\, \frac{\cos(\frac{n}{18}\pi)-\cosh\theta}{2\cos^2(\frac{n}{18}\frac{\pi}{2})}
\eeq
\beq
g_n(\theta)\,=\, \exp\left[2\int_0^\infty
  \frac{\intd x}{x}\frac{\cosh[(n/18-1/2)x]}{\cosh x/2\sinh
  x}\sin^2[(\rmi\pi-\theta)x/2\pi]\right]\ .
\eeq
The $Q_{ii}^{t}(\theta)$ are polynomials in
$\cosh(\theta)$. An upper bound  on the degree $d_P$ of these polynomials can be obtained from the following general formula for the asymptotic
 behaviour of form factors \cite{DM}:
 \begin{equation}
 \lim_{|\theta_i| \to \infty} F^{\varphi}_{a_1,...,a_n}(\theta_1,...,\theta_n)\sim
  \rme^{y_\varphi|\theta_i|},
  \qquad y_\varphi \leq \Delta_\varphi \label{upper}.
 \end{equation}
This relationship fixes $d_P =1$ for  both  $Q_{11}^{t}(\theta)$ and $Q_{22}^{t}(\theta)$:
\beq
Q_{ii}^{t}(\theta)\,=\, a_{ii}^{t} + b_{ii}^{t}\cosh(\theta),\qquad i=1,2\ .
\eeq
These coefficients $a_{ii}^t$ and $b_{ii}^t$ can be 
expressed in terms of the three-particle coupling constants $\Gamma_{11}^2$, $\Gamma_{11}^4$,
$\Gamma_{22}^2$, $\Gamma_{22}^4$ and the one-particle form factors $F_2^{t}$, $F_4^{t}$ by means of the residue equations
\begin{equation}
\label{eq.res2}
-\rmi\lim_{\theta\to \rmi u_{ab}^c} (\theta-\rmi u_{ab}^c)F_{ab}^\varphi
(\theta)=\Gamma_{ab}^cF_c^\varphi
\end{equation}
for the fusions
\begin{equation}
\label{eq.fus1}
A_1\times A_1 \to A_2,\qquad u_{11}^2=\frac{10}{18}\pi,
\end{equation}
\begin{equation}
A_1\times A_1 \to A_4,\qquad u_{11}^4=\frac{2}{18}\pi,
\end{equation}
\begin{equation}
A_2\times A_2 \to A_2,\qquad u_{22}^2=\frac{12}{18}\pi,
\end{equation}
\begin{equation}
\label{eq.fus2}
A_2\times A_2 \to A_4,\qquad u_{22}^4=\frac{8}{18}\pi,
\end{equation}
where $u_{ab}^c$ denotes the fusion angle.
The expressions we found are the following:
\begin{eqnarray}
\label{eq.expr1}
a_{11}^{t} & =1.14107 \cdot \Gamma_{11}^2F_2^{t}-1.77654 \cdot \Gamma_{11}^4F_4^{t}\\
b_{11}^{t} & =-1.21431 \cdot \Gamma_{11}^2F_2^{t}-10.2307 \cdot\Gamma_{11}^4F_4^{t}\\
a_{22}^{t} & =0.306459 \cdot \Gamma_{22}^2F_2^{t}-1.37383 \cdot \Gamma_{22}^4F_4^{t}\\
\label{eq.expr2}
b_{22}^{t} & =-1.76483 \cdot \Gamma_{22}^2F_2^{t}-2.74766 \cdot\Gamma_{22}^4F_4^{t}.
\end{eqnarray}
In terms of these coefficients
\begin{equation}
\label{eq.aaa}
F_{ii}^{t}(\rmi\pi)=a_{ii}^{t}-b_{ii}^{t},
\qquad i=1,2.
\end{equation}
The squares of three-particle coupling constants can be obtained from the S-matrix
elements
\begin{eqnarray}
\label{eq.s1}
S_{11}(\theta) &=-f_{10}(\theta)f_2(\theta)\\
\label{eq.s2}
S_{22}(\theta) &=f_{12}(\theta)f_8(\theta)f_2(\theta),
\end{eqnarray}
where
\beq
f_n(\theta)=\frac{\tanh\frac{1}{2}(\theta+\rmi\pi\frac{n}{18})}{\tanh\frac{1}{2}(\theta-\rmi\pi\frac{n}{18})},
\eeq
by using the residue equation
\begin{equation}
\label{eq.res1}
-\rmi\lim_{\theta\to \rmi u_{ab}^c}
(\theta-\rmi u_{ab}^c)S_{ab}(\theta)=(\Gamma_{ab}^c)^2
\end{equation}
for the fusions (\ref{eq.fus1})-(\ref{eq.fus2}). We obtained
\begin{eqnarray}
\label{eq.gamma1}
(\Gamma_{11}^2)^2 &=4.838705173^2 \qquad
(\Gamma_{11}^4)^2 &=1.225805260^2\\
(\Gamma_{22}^2)^2 &=11.15518618^2 \qquad
(\Gamma_{22}^4)^2 &=19.10015279^2. \label{eq.gamma2}
\end{eqnarray}
Assuming $\Gamma_{ii}^2>0$, $\Gamma_{ii}^4>0$, these results can be substituted
into (\ref{eq.expr1})-(\ref{eq.expr2}), and then (\ref{eq.aaa}) takes the form of (\ref{eq.aa}) and (\ref{eq.bb}).

\appsection
{\bf Calculation of form factors by TCSA}

\noindent
A one-particle form factor of a field $\varphi$ of conformal weights $(\Delta,\Delta)$ at infinite $R$
can be obtained as the limit of the finite
$R$ form factor:
\begin{equation}
\label{eq.limit1}
F_A^{\varphi}=\brakettt{0}{\varphi(0)}{A}=\lim_{R\to\infty}
R^{-2\Delta}\sqrt{Rm}\brakettt{0}{\varphi(0)}{A}_R\ ,
\end{equation}
where the subscript $R$ on the right hand side indicates that the matrix element
should be calculated in the theory defined at size $R$ and $m$ is the mass of
the particle $\ket{A}$. In this formula  $\ket{A}$ is a
zero-momentum state and it
is assumed that $\braket{0}{0}_R=\braket{A}{A}_R=1$.
In the framework of the TCSA the absolute value of the matrix elements
$\brakettt{0}{\varphi(0)}{A}_R$ can be
calculated.
For a general matrix
element the formula
\begin{equation}
\label{eq.me}
\frac{\brakettt{A}{\varphi
    (0)}{B}}{\sqrt{\braket{A}{A}}\sqrt{\braket{B}{B}}}=
\frac{\sum_{a,b}A^*_aB_b\brakettt{a}{\varphi(0)}{b}}{ \sqrt{\sum_{a,b}G_{ab}A_a^*A_b}\sqrt{\sum_{a,b}G_{ab}B_a^*B_b} }
\end{equation}
holds, where the subscript $R$ is suppressed, it is not assumed that the
eigenvectors $\ket{A}$
and $\ket{B}$ are normalized to unity, $A_a$ and $B_b$ are expansion
coefficients with respect to the conformal basis used in the TCSA:
 $\ket{A}=\sum_a A_a\ket{a}$,   $\ket{B}=\sum_b B_b\ket{b}$, and
$G_{ab}=\braket{a}{b}$ is the inner product matrix of the conformal basis
vectors.
 The expansion coefficients
$A_a$ and $B_b$ are calculated numerically up to overall constant factors;
they are provided by the routine that one uses for the diagonalization of the
Hamiltonian operator.

A two-particle form factor $F_{AA}^\varphi(\theta)$ at $\theta=\rmi\pi$
at infinite $R$ can be obtained as the
following limit of finite volume matrix elements:
\begin{eqnarray}
\label{eq.limit2}
\fl
F^{\varphi}_{AA}(\rmi\pi)=
\brakettt{A(0)}{\varphi(0)}{A(0)}\nonumber \\
=\lim_{R\to\infty}
R^{-2\Delta}Rm(\brakettt{A}{\varphi(0)}{A}_R-\brakettt{0}{\varphi(0)}{0}_R).
\end{eqnarray}
It is assumed that $A$ is a self-conjugate particle, and
normalization conditions similar to those for (\ref{eq.limit1}) apply. In the
framework of the TCSA the matrix
elements $\brakettt{A}{\varphi(0)}{A}_R$ and
$\brakettt{0}{\varphi(0)}{0}_R$
can be calculated in a similar way as the matrix element in
(\ref{eq.limit1}), i.e.\ by the formula (\ref{eq.me}).

\appsection
{\bf Calculation of corrections to vacuum energy densities}

\noindent 
The first order correction (\ref{eq.vd}) to the vacuum energy density $\epsilon_{vac} = \lim_{R\to \infty} E_{vac}(R)/R$ at $g_2=1$, $g_4=0$ on the $E_7$-related axis was calculated by the following formula:
\begin{equation}
\delta \epsilon_{vac} =g_4\frac{\intd \epsilon_{vac}}{\intd g_4}(g_4=0) =\lim_{R\to \infty} 2\pi g_4 \lambda_4
R^{-2\Delta_4}\brakettt{0}{t(0)}{0}_R\ .
\end{equation}

The numerical calculation of the first order corrections  to the
energy densities of the $\ket{+}$, $\ket{-}$, $\ket{0}$ vacua
existing at the first order phase  transition point is slightly
more complicated, since these vacua are degenerate in infinite
volume ($R=\infty$). Because of this degeneracy, one has to use,
in general, degenerate perturbation theory, i.e.\ one  obtains the
first-order corrections by calculating the eigenvalues of the $3$
by $3$ matrix constituted by the $9$
 matrix
elements $\brakettt{+,-,0}{H_{p}}{+,-,0}$  of the perturbing
operator $H_p$ between the three vacua. Those combinations of the
vacua which evolve into the split vacua existing at nonzero
perturbation are given by the eigenvectors of this matrix.

 More precisely, in our case the first
order corrections to the  energy
densities of the vacua are given by the eigenvalues
of the matrix
\beq
\label{eq.M}
M_{ij}=\lim_{R\to \infty} 2\pi g_2 \lambda_2
R^{-2\Delta_2}\brakettt{U_i}{\varepsilon(0)}{U_j}_R,
\eeq
where $i,j=1,2,3$,  $g_4$ has the value $-1$,
and $\ket{U_{1}}_R$,  $\ket{U_{2}}_R$,  $\ket{U_{3}}_R$
 are the three vacua at cylinder circumference $R$,
normalized to $1$. The numbers
$\brakettt{U_i}{\varepsilon(0)}{U_j}_R$ can be calculated as described above
(see equation (\ref{eq.me})). It is important, in general,
that $\ket{U_{1}}_R$,  $\ket{U_{2}}_R$,  $\ket{U_{3}}_R$ should
be orthogonal, otherwise the inverse of their inner product matrix has to be
included in (\ref{eq.M}).
In our case the vacua  are split at finite $R$ by tunneling effects,
so $\ket{U_{1}}_R$,  $\ket{U_{2}}_R$,  $\ket{U_{3}}_R$ belong to different
eigenvalues, therefore they can be identified
uniquely and  they are orthogonal.

The problem of the diagonalization of $\brakettt{U_i}{\varepsilon(0)}{U_j}_R$  can be simplified by symmetry
considerations in our case: two of the three vacua $\ket{U_{1}}_R$,  $\ket{U_{2}}_R$,  $\ket{U_{3}}_R$ lie in
the spin reversal even subspace, the third one lies in the odd subspace. We
choose $\ket{U_1}_R$ and $\ket{U_2}_R$ to be the even vacua and $\ket{U_3}_R$
to be the odd one.
Due to the spin reversal symmetry of
  the $\varepsilon$ perturbation, the matrix elements
$\brakettt{U_{1,2}}{\varepsilon(0)}{U_3}_R$ are all $0$,
therefore the diagonalization problem can be treated separately in the even
and odd sectors.
The odd vacuum  (which becomes
$\frac{\ket{+}-\ket{-}}{\sqrt{2}}$ in infinite volume) thus evolves into the
 odd (possibly metastable)  vacuum of the perturbed theory,  and
the first-order
    correction to its energy density is given by
\beq
\delta \epsilon_{U_3}=\lim_{R\to \infty} 2\pi g_2 \lambda_2
R^{-2\Delta_2}\brakettt{U_3}{\varepsilon(0)}{U_3}_R.
\eeq
In the even sector, one has to consider the $2$ by $2$ matrix
$\brakettt{U_{1,2}}{\varepsilon(0)}{U_{1,2}}_R$. Since the duality $D$ maps the
even sector into itself and commutes with the unperturbed Hamiltonian operator
in this sector, and $\ket{U_1}_R$ and $\ket{U_2}_R$ have
different energies, $\ket{U_1}_R$ and $\ket{U_2}_R$ have to be eigenstates of
$D$. The eigenvalues can be $+1$ or $-1$, because $D^2=1$. We also know that
$\varepsilon$ changes sign under duality, therefore
$\brakettt{U_1}{\varepsilon(0)}{U_1}_R=\brakettt{U_2}{\varepsilon(0)}{U_2}_R=0$.
$\varepsilon$ is also self-adjoint, so
$\brakettt{U_1}{\varepsilon(0)}{U_2}_R=\brakettt{U_2}{\varepsilon(0)}{U_1}_R^*$.
This number can be nonzero if the eigenvalues of  $\ket{U_1}_R$ and
$\ket{U_2}_R$ with respect to $D$ have opposite sign.
We denote the  eigenvectors of the matrix $\brakettt{U_{1,2}}{\varepsilon(0)}{U_{1,2}}_R$
by  $\ket{V_+}_R$ and $\ket{V_-}_R$.
The eigenvalues of the matrix
$\brakettt{U_{1,2}}{\varepsilon(0)}{U_{1,2}}_R$ are $\pm
|\brakettt{U_{1}}{\varepsilon(0)}{U_{2}}_R|$,  so the correction to the energy
densities of the two vacua $\ket{V_+}_\infty$ and $\ket{V_-}_\infty$ are
\beq
\delta \epsilon_{V_+}=\lim_{R\to \infty} 2\pi g_2 \lambda_2
R^{-2\Delta_2}|\brakettt{U_1}{\varepsilon(0)}{U_2}_R|,
\eeq
and
\beq
\delta \epsilon_{V_-}=\lim_{R\to \infty} -2\pi g_2 \lambda_2
R^{-2\Delta_2}|\brakettt{U_1}{\varepsilon(0)}{U_2}_R|.
\eeq
Here  we have chosen
 $\ket{V_+}_R$ to have  the positive eigenvalue and  $\ket{V_-}_R$ to have
the negative one.
Note that these formulas would not be meaningful without taking absolute
values on the right hand side, since  $\ket{U_1}_R$ and
$\ket{U_2}_R$ are defined up to arbitrary phase factors.

The analytic result (\ref{eq.ftv}) described below, the  Landau-Ginzburg picture and
our numerical calculations of energy spectra all  show that one vacuum gets a
negative correction and two vacua get the same positive correction as we move
into the high temperature phase. It follows then that
\beq
\label{eq.uvv}
\delta\epsilon _{U_3}=
\delta\epsilon _{V_+}=-\delta\epsilon _{V_-},
\eeq
and $\Delta E=2\delta\epsilon_{U_3}$. The direct TCSA calculation of
$\delta\epsilon _{U_3}$ and $\delta\epsilon _{V_+}$ confirms the equality
$\delta\epsilon _{U_3}=
\delta\epsilon _{V_+}$ and yields the result (\ref{eq.deltae}) for $\Delta E$.

Having discussed the calculation of the energy density corrections,
 we turn to the description of the relation between the vacua
 $\ket{U_{1}}_\infty$,  $\ket{U_{2}}_\infty$,  $\ket{U_{3}}_\infty$;
 $\ket{V_+}_\infty$,
$\ket{V_-}_\infty$ and
 $\ket{+}$,
$\ket{-}$, $\ket{0}$.
Equation (\ref{eq.uvv}) implies that
\beq
\label{eq.u1}
\brakettt{U_3}{\varepsilon}{U_3}_\infty =
|\brakettt{U_1}{\varepsilon}{U_2}_\infty|,
\eeq
and as we mentioned above, we also have
\beq
\label{eq.u2}
\brakettt{U_1}{\varepsilon}{U_2}_\infty =
\brakettt{U_2}{\varepsilon}{U_1}_\infty^*,
\eeq
and all other matrix elements of $\varepsilon$ between $\ket{U_{1}}_\infty$,
$\ket{U_{2}}_\infty$,  $\ket{U_{3}}_\infty$
are zero. $\brakettt{U_1}{\varepsilon}{U_2}_\infty$ can be made real and
positive by a phase redefinition of
$\ket{U_{1}}_\infty$ or  $\ket{U_{2}}_\infty$.

The matrix elements of $\varepsilon$ between the $R=\infty$ eigenvectors $\ket{+}$,
$\ket{-}$, $\ket{0}$ were calculated exactly in \cite{FLZZ,FMS}:
\beq
\label{eq.ftv}
\brakettt{+}{\varepsilon}{+}=\brakettt{-}{\varepsilon}{-}=-\brakettt{0}{\varepsilon}{0}=\Lambda_2>0,
\eeq
and the nondiagonal matrix elements are $0$ (the numerical value of $\Lambda_2$ is
given by (\ref{eq.B2}), but it is irrelevant here).
This result, together with the spin reversal properties, (\ref{eq.u1}) and
(\ref{eq.u2}),  allows us to identify
 $\ket{U_{1}}_\infty$,  $\ket{U_{2}}_\infty$,  $\ket{U_{3}}_\infty$ in terms
of
 $\ket{+}$,
$\ket{-}$, $\ket{0}$ as follows:
\begin{eqnarray}
\label{eq.uu1}
\ket{U_1}_\infty & = & \alpha\,  \frac{\ket{+}+\ket{-}}{2}+ \beta\, \frac{\ket{0}}{\sqrt{2}}\\
\label{eq.uu2}
\ket{U_2}_\infty & = &  \alpha\,  \frac{\ket{+}+\ket{-}}{2}- \beta\, \frac{\ket{0}}{\sqrt{2}}\\
\ket{U_3}_\infty & = & \gamma\, \frac{\ket{+}-\ket{-}}{\sqrt{2}},
\end{eqnarray}
where it is also assumed that the phases of   $\ket{U_{1}}_\infty$ and
 $\ket{U_{2}}_\infty$ are chosen so that
 $\brakettt{U_1}{\varepsilon}{U_2}_\infty$ is real and positive. $\alpha$,
 $\beta$ and $\gamma$ are unknown phase factors which
are not determined by the information we have described so far, and they can
 be set to $1$ by the following redefinition of the phases of
 $\ket{+}$, $\ket{-}$, $\ket{0}$ and $\ket{U_3}_R$:
$\ket{+}\to \frac{1}{\alpha} \ket{+}$, $\ket{-}\to \frac{1}{\alpha} \ket{-}$,
$\ket{0} \to \frac{1}{\beta}\ket{0}$,  $\ket{U_3}_R \to
 \frac{\gamma}{\alpha} \ket{U_3}_R$.
These redefinitions leave the previously stated  relations involving these
 vectors unchanged.
 $\ket{V_+}_\infty$ and
$\ket{V_-}_\infty$ can now be expressed as
\begin{eqnarray}
\ket{V_+}_\infty  & = \frac{\ket{U_1}_\infty+\ket{U_2}_\infty}{\sqrt{2}}
  & =\  \frac{\ket{+}+\ket{-}}{\sqrt{2}}\\
 \ket{V_-}_\infty  & = \frac{\ket{U_1}_\infty-\ket{U_2}_\infty}{\sqrt{2}}
 & = \ket{0}.
\end{eqnarray}
We note finally that
since  $\ket{U_{1}}_\infty$ and  $\ket{U_{2}}_\infty$ are duality eigenstates
with eigenvalues of opposite sign, equations (\ref{eq.uu1}) and (\ref{eq.uu2})
imply that the duality maps $\ket{0}$ and $\frac{\ket{+}+\ket{-}}{\sqrt{2}}$
into each other.

\appsection
{\bf Kink form factors}

\noindent
Along the first-order phase transition line ($g_2=0$, $g_4<0$), when the TIM has three degenerate vacua and an exact supersymmetry, in the infinite volume ($R= \infty$) the two-kink form factors
\begin{eqnarray*}
 \brakettt{0}{\varepsilon(0)}{K_{0-}(\theta_1)K_{-0}(\theta_2)} & \equiv
F_{0-}^\varepsilon (\theta_1-\theta_2)\ ,\\
 \brakettt{0}{\varepsilon(0)}{K_{0+}(\theta_1)K_{+0}(\theta_2)} & \equiv
F_{0+}^\varepsilon (\theta_1-\theta_2)\ , \\
 \brakettt{-}{\varepsilon(0)}{K_{-0}(\theta_1)K_{0-}(\theta_2)} & \equiv
F_{-0}^\varepsilon (\theta_1-\theta_2)\ ,\\
 \brakettt{+}{\varepsilon(0)}{K_{+0}(\theta_1)K_{0+}(\theta_2)} & \equiv
F_{+0}^\varepsilon (\theta_1-\theta_2)
\end{eqnarray*}
are  the following \cite{D}:
\[
F_{0-}^\varepsilon (\theta)=F_{0+}^\varepsilon (\theta)\ ,\qquad
F_{-0}^\varepsilon (\theta)=F_{+0}^\varepsilon (\theta)\ ,
\]
\begin{equation}
F^{\varepsilon}_{0-}(\theta)=-\rmi (U_{-}-U_0)\frac{\rme^{-\frac{\gamma}{2}(\pi+\rmi\theta)}}{p\sinh\frac{1}{p}(\theta-\rmi
\pi)}F_0(\theta)\ , \label{eq.ff1}
\end{equation}
\beq
F^{\varepsilon}_{-0}(\theta)=\rmi (U_{-}-U_0)\frac{\rme^{\frac{\gamma}{2}(\pi+\rmi\theta)}}{p\sinh\frac{1}{p}(\theta-\rmi
\pi)}F_0(\theta)\ , \label{eq.ff2}
\eeq
where
\[
U_0=\brakettt{0}{\varepsilon}{0}\ ,\qquad
U_{-}=\brakettt{-}{\varepsilon}{-}=\brakettt{+}{\varepsilon}{+}\ ,
\]
\begin{equation}
F_0(\theta)=-\rmi\sinh\frac{\theta}{2}\,\exp\left[\int_0^\infty\frac{\intd x}{x}\,
\frac{\sinh(1-p)\frac{x}{2}}{\sinh\frac{px}{2}\cosh\frac{x}{2}}\,
\frac{\sin^2(\rmi\pi-\theta)\frac{x}{2\pi}}{\sinh x}\right], \nonumber
\label{eq.fo}
\end{equation}
\[
 \gamma=\frac{1}{2\pi}\ln 2\ ,\qquad
p=4\ .
\]
These form factors have a pole at $\rmi\pi$, as  can be explicitly seen from
 (\ref{eq.ff1}),  (\ref{eq.ff2}).

\end{document}